# Enhanced Superconductivity and Mixed-dimensional Behaviour in Infinite-layer Samarium Nickelate Thin Films

Mingwei Yang[1,2,12], Heng Wang[3,12], Jiayin Tang[1,12], Junping Luo[4], Xianfeng Wu[3], Wenjing Xu[1], Aile Wang[5,6], Yuetong Wu[7], Ruilin Mao[4], Ze Wang[8], Zhicheng Pei[1], Guangdi Zhou[3], Zhengang Dong[1,2], Bohan Feng[1,2], Lingchi Shi[1], Wenjie Meng[8], Chuanying Xi[8], Li Pi[8], Qingyou Lu[5,6,8], Jun Okamoto[9], Hsiao-Yu Huang[9], Di-Jing Huang[9], Haoliang Huang[3,10], Qisi Wang[7], Peng Gao[4,11], Zhuoyu Chen[3,10*] and Danfeng Li[1,2*]

[1]*Department of Physics, City University of Hong Kong, Kowloon, Hong Kong SAR, China.*
[2]*Shenzhen Research Institute of City University of Hong Kong, Shenzhen 518057, China.*
[3]*Department of Physics, State Key Laboratory of Quantum Functional Materials, and Guangdong Basic Research Center of Excellence for Quantum Science, Southern University of Science and Technology, Shenzhen 518055, China.*
[4]*International Center for Quantum Materials, and Electron Microscopy Laboratory, School of Physics, Peking University, Beijing 100871, China.*
[5]*Hefei National Research Center for Physics Sciences at the Microscale, University of Science and Technology of China; Hefei 230026, China.*
[6]*Hefei National Laboratory, University of Science and Technology of China; Hefei 230026, China.*
[7]*Department of Physics, The Chinese University of Hong Kong, Shatin, Hong Kong SAR, China.*
[8]*Anhui Key Laboratory of Low-Energy Quantum Materials and Devices, High Magnetic Field Laboratory, HFIPS, Chinese Academy of Sciences, Hefei 230031, China.*
[9]*National Synchrotron Radiation Research Center, Hsinchu 30076, Taiwan.*
[10]*Quantum Science Center of Guangdong-Hong Kong-Macao Greater Bay Area, Shenzhen 518045, China.*
[11]*Collaborative Innovation Center of Quantum Matter, Beijing 100871, China.*
[12]*These authors contributed equally*: Mingwei Yang, Heng Wang, Jiayin Tang.
[*]*Corresponding author*: chenzhuoyu@sustech.edu.cn, danfeng.li@cityu.edu.hk.

**Abstract**

Rare-earth infinite-layer nickelates represent an emerging class of unconventional superconductors, with materials synthesis largely limited to early lanthanide compounds. Here, we report the synthesis and characterization of phase-pure superconducting samarium-based infinite-layer nickelate thin films, including the first demonstration of $Sm_{1-x}Sr_xNiO_2$, along with co-doped variants incorporating europium and calcium. These films, grown on LSAT (001) substrates, exhibit coherent lattice structures up to ~ 9 nm thickness with minimal stacking faults. The co-doped compounds achieve a record-small *c*-axis parameter of 3.26 Å and display remarkable superconducting transition temperatures up to 32.5 K. These results establish a clear correlation between decreasing *c*-axis parameter and increasing critical temperature across different rare-earth systems. In addition, angle-dependent magnetoresistance investigations reveal the existence of a hybrid mixture of 2D and 3D superconductivity in this novel system with enhanced coupling between the rare-earth 5*d* and Ni 3*d* orbitals, confirmed by resonant inelastic X-ray scattering experiments. As the concentration of Eu increases, the system exhibits a clear tendency towards 3D superconductivity. Furthermore, we observe distinctive negative magnetoresistance even in the superconducting state in the europium-containing samples. These findings advocate clear materials design principles for higher transition temperatures and exotic physics in infinite-layer nickelate superconductors through structural engineering of the rare-earth site.

## Introduction

Superconducting infinite-layer (IL) nickelates, since discovered[1] stand as a subject of continuous research interest[2,3], featured by their potential of being an analogous model system to high-$T_c$ cuprates[4] as well as by the essential reductive materials synthesis pathways to the unconventional Ni valence state[5]. Their electronic structure displays a few salient features as compared to cuprates, including the unusual role of the rare-earth orbitals[6–8] and the belated multi-band nature[9–11] with the presence of a 'self-doping' effect[8,12] and clear electron pockets at the Fermi surface[9,10]. This raises an interesting question regarding the role of the detailed configuration of the *f* electrons of the rare-earth cations and their affiliated local moment[13,14], and the possible interaction with the itinerant charges of 3*d* character[15], a hall mark of heavy-fermion physics. However, due to the significant growth challenges[16], materials synthesis efforts have largely been limited to the La-[17–19], Pr-[20,21] and Nd-series[22–26]: many of the anticipated 'Kondo-like' physics[12,27,28] remain to be seen. Besides, given the discrepancies reported on the (an-)isotropic nature of different rare-earth IL nickelate systems[15], designing and preparing additional IL nickelates with tuneable interplanar spacings and detailed electronic configurations, and to investigate the dimensionality of their superconducting ground state are highly demanded.

Another motivation for extending the materials paradigm to late-lanthanide IL nickelates is to drive higher $T_c$. Indeed, a widely employed strategy for tuning $T_c$ is to exert chemical pressure by utilising elements of smaller cationic radius[29,30]. Such evolution of the structural tolerance parameter, enabled by the rare-earth sublattices, directly modulates the long-range chemical bonding configurations embedded in the lattice distortion, resulting in direct control of the electronic bandwidth and correlation strength[31–34]. This has produced various competing and often intertwined emerging orders of both commensurate and incommensurate nature, which drastically correlate to local charge and/or spin instabilities, as well as superconductivity[4,35]. In this regard, example systems with properties largely underpinned by such rare-earth dependence span across from famous cuprates[4] and pnictides[36], to intermetallic compounds[37], to the Ruddlesden-Popper (RP) nickelate of various nickel valence[38–40], the latter of which has been the focus of recent investigations[41,42].

For superconducting IL nickelates, hydrostatic high-pressure experiments indicate that a shrinking lattice parameter may contribute to a higher $T_c$[21]. Furthermore, early work on Nd- and Pr-based IL nickelates on substrates of smaller lattice constants, such as $(LaAlO_3)_{0.3}(Sr_2TaAlO_6)_{0.7}$ (LSAT) and $NdGaO_3$ (NGO) corroboratively suggest an enhancement of $T_c$ (for instance, for $Nd_{0.8}Sr_{0.2}NiO_2$, the onset $T_c$ is ~ 15 K on the $SrTiO_3$ (001) substrate[22,23], ~ 19.3 K on the LSAT (001) substrate[24], and ~ 25.7 K on the $NdGaO_3$ (110) substrate[25]), in which the epitaxial strain may be instrumental[43], similar to that in the doped cuprate thin films[44]. All these aspects highlight the intricate yet useful role of the smaller lattice spacing in IL nickelates for higher $T_c$, which warrants further investigations, particularly through introducing smaller rare-earth ions. In addition, discrepancies in magneto-transport behaviours, dimensionality, and the pertained magnetic footprints across La-, Pr-, and Nd-series IL nickelates, which stem from the variations in the spin configuration of the A-site ions[19–26,45], call for the quest for new members of the IL nickelate family.

With this notion and motivated by the recent observation of a $T_c$ above 35 K in infinite-layer $Sm_{1-x-y-z}Ca_xSr_yEu_zNiO_2$ nickelates on $NdGaO_3$ (110) substrates[46], we have synthesised phase-pure Sm-based IL nickelate thin films on LSAT (001) substrates and carefully studied their superconducting transport properties and electronic structure. The highest $T_c$ of 32.5 K has been observed. Atomically-resolved scanning transmission electron microscopy (STEM) images reveal the high quality of our films of up to ~ 9 nm with minimal stacking faults. In addition, driven by the fact that a compositional A-site and the distribution in cationic radius may also significantly affect superconductivity[47,48], with the emergence of ordered phases (spin and/or charge orderings)[4,49,50], we furthered our materials growth efforts to the synthesis of a combinational series of Sm-based superconducting IL nickelate thin films with A-site (in $ABO_2$) composition of Eu, Sr, and Ca. In particular, superconducting $Sm_{1-x}Sr_xNiO_2$ have been demonstrated for the first time. We observed a generic mixed nature of both two- and three-dimensional superconductivity revealed by high-magnetic-field measurements across all samples. As the concentration of Eu increases, the system exhibits a tendency towards 3D superconductivity. This behaviour is in concert with an enhanced coupling of the rare-earth 5*d* and Ni 3*d* orbitals, corroborated by the resonant X-ray scattering measurements. Our results add new superconducting IL nickelates with high $T_c$ into this intriguing materials family and underscore a fundamental trend of an increasing $T_c$ value as the *c*-axis constant shrinks. These findings contribute to the understanding of the superconductivity mechanism, offering clear guidelines to uncovering additional IL nickelate superconductors with even higher $T_c$.

**Results**

Multiple sets of Sm-based IL nickelate films of thicknesses of ~ 10 nm with various compositions and a $SrTiO_3$ capping layer of ~ 2 nm were prepared using pulsed laser deposition (PLD). We note a narrower growth parameter window for high-quality, hole-doped Sm nickelate as compared to that of Nd version. Details on the sample synthesis can be found in Methods.

The precursor perovskite phases ($Sm_{0.8}Sr_{0.2}NiO_3$, $Sm_{0.74}Ca_{0.01}Sr_{0.19}Eu_{0.06}NiO_3$, $Sm_{0.79}Ca_{0.04}Sr_{0.05}Eu_{0.12}NiO_3$ and $Sm_{0.75}Ca_{0.05}Eu_{0.2}NiO_3$, as well as others with different Ca concentrations) were grown on LSAT (001) substrates. The detailed characterizations are shown in the Supplementary Information (Figure S1). The presence of prominent (001) reflections[16] and finite-size fringes around the main film peaks in the X-ray diffraction (XRD) scans (Figure S1a) suggests a good film quality. The *c*-axis lattice parameters calculated from the data are 3.75 Å, 3.74 Å, 3.74 Å and 3.729 Å, respectively. These values are smaller than that in $Nd_{1-x}Sr_xNiO_3$[51], in line with a smaller tolerance factor. The corresponding resistivity curves $\rho(T)$ of the first three samples are shown in Figure S1b. Unlike previously reported[46], all these samples show metallic behaviour down to low temperature (2 K), consistent with charge-induced suppression of the metal-insulator transition[51,52]. These intrinsic metallic behaviours inform the high crystallinity of the precursor phase. We note a few anomalies of the slope change in the $\rho(T)$ curve for $Sm_{0.79}Ca_{0.04}Sr_{0.05}Eu_{0.12}NiO_3$ that can be attributed to the magnetism in the system. The local 'plateau' in resistivity at ~ 60 K may correspond to an antiferromagnetic ordering temperature of Sm, while the local resistivity minimum at ~ 8 K correlates with a spin-disorder-related metal-insulator transition induced by a partial substitution of Sm with Eu[53]. These transitions are more revealing on a derivative resistivity curve (see Figure S2 in the Supplementary Information).

The reduction to the IL phase was done using an in-situ setup (see Methods for details and Figure S4 for the reduction process). Figure 1 (a-d) illustrates the XRD scans of the films in the IL phase upon reduction: $Sm_{0.8}Sr_{0.2}NiO_2$, $Sm_{0.74}Ca_{0.01}Sr_{0.19}Eu_{0.06}NiO_2$, $Sm_{0.79}Ca_{0.04}Sr_{0.05}Eu_{0.12}NiO_2$, and $Sm_{0.75}Ca_{0.05}Eu_{0.2}NiO_2$ (for simplicity and consistency, we name them as SSNO, $SCSE_{0.06}$, $SCSE_{0.12}$ and $SCE_{0.2}$, respectively), all of which show significant (00l) film peaks, indicative of high quality, corresponding to the *c*-axis lattice constant of 3.307 Å, 3.296 Å, 3.26 Å and 3.273 Å, respectively. The $SCSE_{0.12}$ represents the smallest *c*-axis parameter so far reported for the IL nickelate thin films, with 12 % Eu and 4 % Ca introduced to partially replace Sm and Sr, which have larger ionic radius[54]. These samples all show superconductivity at low temperatures with a $T_c$ onset of ~ 15 – 32.5 K (Figure 1, e-h). In particular, superconductivity in $Sm_{1-x}Sr_xNiO_2$ was observed for the first time. All four samples do not show clear *T*-linear dependent behaviour, implying the doping level may not be at the optimal level[24]. For $SCE_{0.2}$, it is worth noting that we initially aimed at achieving single-phase superconducting $Sm_{0.8}Eu_{0.2}NiO_2$; however, we noticed that the introduction of Ca is of crucial importance for obtaining high-quality IL nickelates under the current optimised reduction conditions, as illustrated in Figure S3. Therefore, we have selected $SCE_{0.2}$ for our study here.

The high quality of the samples is confirmed by STEM. Figure 1 (i-j) displays cross-sectional high-angle annular dark field (HAADF) STEM images of an SSNO sample and an $SCSE_{0.06}$ sample. From the images, despite occasional RP-type extended defects, single-phase IL nickelate layers of ~ 9 nm with homogeneous lattice coherency and sharp interfaces with the LSAT substrates can be clearly seen. The lattice spacings measured from the atomically resolved HAADF images are ~ 3.34 Å, ~ 3.28 Å for the SSNO and $SCSE_{0.06}$ samples, generally consistent with the XRD results.

Magnetotransport measurements were performed in-house under magnetic fields up to 14 T perpendicular and parallel to the $NiO_2$ planes for the samples (Figure 2). The Meissner diamagnetic responses were recorded below a characteristic temperature $T_M$ of 5.4 K, 7.2 K, 4.6 K and 15.8 K for SSNO, $SCSE_{0.06}$, $SCSE_{0.12}$, and $SCE_{0.2}$, respectively, in the mutual inductance measurements, with the driving and pickup coils vertically aligned above and below the sample (see Methods). The data is displayed in the insets of Figure 2 (e-h). These $T_M$ values are slightly lower than zero-resistance $T_c$ (defined as $T_{c,0}$), consistent with theoretical predictions for layered superconductors, where the flux penetration dynamics (governed by weak interlayer coupling) result in delayed emergence of complete diamagnetic shielding compared to resistive transitions[55]. The resistivity data sets show a clear anisotropic superconducting characteristic (across Figure 2, a-h and more visual in Figure 2, i-l) independent of $T_c$ value ($T_{c,50\%}$, defined as the midpoint of the resistive transition, $\rho(T_c) = \rho_{50\%}$, where $\rho_{50\%}$ is the 50% of the normal-state resistivity; see Figure S5 in the Supplementary Information for definition), and can be largely captured by the Tinkham's framework for 2D superconductors despite the requirement of the thickness ($d$) being smaller than the Ginzburg-Landau (GL) coherence length ($\xi_{GL}$) at $T = 0$. Extrapolation of a linear-$T$ dependence of the upper critical field ($H_{c2}$) close to $T_c$ in the out-of-plane field ($H//c$) configuration (Figure 2, i-l) gives $\xi_{GL}$ at $T = 0$ to be 3.3 nm, 2.8 nm, 2.1 nm, 1.7 nm for SSNO, $SCSE_{0.06}$, $SCSE_{0.12}$ and $SCE_{0.2}$, respectively. The $H_{c2}(t) \sim (1-t)^{1/2}$ (where $t = T/T_c$ is the reduced temperature) behaviour close to $T_c$ under in-plane field ($H \perp c$) for SSNO and $SCSE_{0.06}$, as shown in Figures 2i and 2j, is attributed to the dominant spin paramagnetic pairing breaking[56,57]. For $SCSE_{0.12}$ and $SCE_{0.2}$, a different behaviour is observed and will be discussed later. The generic anisotropy of the system across all samples can also be corroborated by an anisotropic field response of vortex motion: as illustrated in Figure 2 (m-p), the field-dependent vortex activation energy, $U_0 \sim H^{-\alpha}$, extracted from the $T$-dependent Arrhenius relationship based on the thermally-activated flux flow (TAFF) model[58]. The data show a large difference of the power-law exponent $\alpha$ for the two field directions, indicating a strong anisotropy in vortex pinning strength in the Sm-based system. Figure 2 (n-p) reveals a monotonic decrease in $\alpha$ values for $H \perp c$ with increasing Eu concentration, indicative of a progressive strengthening of the interlayer coupling strength.

The samples with Eu as an effective component show interesting response to magnetic field: $SCSE_{0.06}$, $SCSE_{0.12}$ and $SCE_{0.2}$ all display a sizable negative magnetoresistance in their normal state (insets of Figure 2, b-d for the perpendicular-field case, and Figure S8b for the parallel-field case). Another thing to note is that both $SCSE_{0.12}$ and $SCE_{0.2}$ exhibit a weak but clear 'field-enhanced superconductivity' when $H \perp c$ (Figure 2, g and k, and Figure 2, h and i, respectively). This phenomenon may suggest a coupling between the local moment of the Eu $f$ orbitals and the itinerant Ni 3$d$ electrons.

To access the intrinsic $H_{c2}$ and compare with other IL nickelates, high-magnetic-field measurements (see Methods) were performed on SSNO. The data (Figure S7) shows a clear anisotropic behaviour in $H_{c2}(T)$, which is more reminiscent of that for La- and Pr-based IL nickelates[45,56,59], distinct from the Nd series[26,57]. Indeed, when normalized against $T_c$ at 0 T (defined as $T_{c0}$) and plotted in comparison with data sets of all other rare-earth series (Figure 3, a and b), the $H_{c2}$ - $T$ dependence of the SSNO sample shows a clear violation of the Pauli limit (dash lines) and a strong deviation from the Nd-based IL nickelates, despite that the ground state of $Sm^{3+}$ is a Kramer's doublet with a $4f^5$ electron configuration and a possible enhanced magnetic permeability[15], which is similar to $Nd^{3+}$. This observation further raises questions on the role that the rare-earth ions (in particular, the $4f$ electron configuration) may play in the unconventional nature of superconductivity in IL nickelates: a delicate balance between the detailed magnetic structure and the lattice exerted by chemical pressure.

The SSNO sample was further put on a rotator under the magnetic field for angle-dependent magnetoresistance ($H_{c2}$ versus $\theta$) measurements at $T$ = 5 K. For $H//c$ ($\theta$ = 0°), the magnetic field was large enough to suppress superconductivity: the normal-state can be attained, from which a $H_{c2,50\%}$ can be extracted using 50% of the linear fit to the normal-state magnetoresistance (see Methods for details). The $H_{c2,50\%}$ - $\theta$ data is shown in Figure 3c and can be described by a combination of the Tinkham's model for 2D superconductors and the anisotropic GL model for 3D superconductors using the following formula:

$$\left[\frac{H_{c2}(\theta)\sin(\theta)}{H_{c2}^{H\perp c}}\right]^2 + \beta \times \left[\frac{H_{c2}(\theta)\cos(\theta)}{H_{c2}^{H\parallel c}}\right]^2 + (1-\beta) \times \left|\frac{H_{c2}(\theta)\cos(\theta)}{H_{c2}^{H\parallel c}}\right| = 1 \qquad (1)$$

where $\beta$ is a dimension-less coefficient that informs how '3D' the superconductor behaves, and $\theta$ is the angle of the magnetic field with respect to the normal of the sample surface, as illustrated in the inset of Figure 3c. The best fit (red line) to the data yields a $\beta$ of 0.366, indicating a mixed '2D' and '3D' character.

This behaviour is further validated by the $T_{c,50\%}$ - $\theta$ experiments, which can be performed by measuring $T_{c,50\%}$ from $\rho(T)$ at variant $\theta$ values under a constant in-house magnetic field(at 9 T), as shown in Figure 3d for SSNO. Here, the fitting was done in a similar manner, proportioning both the '2D' and '3D' contributions:

$$T_c(\theta) = T_{c0} + \beta' \times \left\{H_0/(\partial H_c^{H\parallel c}/\partial T) \times [\cos^2(\theta) + \gamma \sin^2(\theta)]^{1/2}\right\}$$
$$+(1-\beta') \times [(T_c^{H\parallel c} - T_{c0})\cos(\theta) + (T_c^{H\perp c} - T_{c0})\sin^2(\theta)] \qquad (2)$$

where $\beta$' is similarly defined as above and $\gamma$ represents the anisotropic mass ratio of the in-plane and out-of-plane electron motion. Note that the second and the last terms on the right-hand side of the formula correspond to the angular dependence of $T_c$ given by the anisotropic GL 3D model and the Tinkham's 2D model, respectively. Again, SSNO shows a clear mixed feature.

Following the same method, $SCSE_{0.06}$, $SCSE_{0.12}$ and $SCE_{0.2}$ were all measured, and their $T_{c,50\%}$ - $\theta$ data, which are shown in Figure 3, e-g, respectively, suggest a mixed '2D + 3D' behaviour for nearly all samples. The overall increase in $\beta$' with higher Eu doping indicates a tendency toward a 3D superconducting ground state of the system.

To further interrogate the possible implications of this '3D' feature, at the electronic structure level, we performed Ni $L_3$-edge resonant inelastic X-ray scattering (RIXS) measurements on an $SCE_{0.2}$ sample (Figure 3, h-j). Consistent with earlier studies on other IL nickelate systems, the *d-d* excitations are observed in the energy-loss range of ~ 1 to 3 eV[6,43,60,61]. The excitation at ~ 0.6 eV points to a sizable hybridization between the Ni $3d_{z^2}$ and the rare-earth $5d_{z^2}$, $5d_{xy}$ states[6,8]. The significantly enhanced intensity under the X-ray beam of $\pi$ polarization at grazing-incidence geometry is consistent with a predominantly out-of-plane orbital symmetry. This hybridization feature appears to be more intense[6,60,61], indicating that the reduced ionic radius of Sm and Eu effectively affects the electronic structure. The resulting enhancement in the electronic bandwidth may be crucial to the enhanced $T_c$.

The negative normal-state Hall coefficient ($R_H$) of the $SCSE_{0.12}$ and $SCE_{0.2}$ samples display no sign change (Figure 4a) down to low temperature, despite a higher total dopant concentration ($x+y+z$ = 0.21 for $SCSE_{0.12}$ and $x+y$ = 0.25 for $SCE_{0.2}$) as compared to that of SSNO ($x+y+z$ = 0.2), suggesting a lower effective hole-doping level. Such effect can also be seen for the $SCSE_{0.06}$ sample ($x+y+z$ = 0.26), where the crossover temperature ($T_{cross}$) for the sign change in $R_H$, is not far from that of SSNO. These are in line with the co-existence of $Eu^{2+}$ and $Eu^{3+}$ states and a shift of the superconducting 'dome' towards higher doping in Eu-doped infinite-layer compounds[26]. In Figure 4b, we further plot $T_c$, $T_{cross}$ as a function of an estimated effective hole doping level (see Figure S9 in Supplementary Information for more analysis) and overlay them with the data measured on high-quality $Nd_{1-x}Sr_xNiO_2$ on LSAT[24]. It can be clearly seen that, upon estimating the fraction of $Eu^{2+}$ and the re-adjustment on doping level, our data follow a similar trend (within error bars).

Last, we summarize $T_c$ versus *c*-axis parameter of our samples grown on LSAT substrates (red rhombus-shaped points at the top-right part) together with data points from previous studies across different systems, illustrated in Figure 4c. These data points are largely located within the yellow-shaded region, for which a clear correlation between $T_c$ and *c*-axis constant reveals the lattice motif (i.e. distance between Ni-O planes) to superconductivity and suggests the key role of the interplane coupling in mediating $T_c$. As indicated in previous studies[21,62], there is a general enhancement of superconductivity as the *c*-axis lattice constant shrinks: we can see from the figure that as the average *c*-axis parameter decreases from ~ 3.43 Å for $La_{1-x}Sr_xNiO_2$ to ~ 3.28 Å for Sm-based compounds, $T_c$ steadily increases from ~ 10 – 12 K to 24 – 28 K or to even above 32 K, making the infinite-layer nickelates towards a 'high-$T_c$' system under ambient pressure[21,46,63].

**Discussions**

Our study was motivated by the recent report on reaching a remarkably high $T_c$ in Sm-based IL nickelate thin films[46]. Our results suggest that the overall smaller A-site ionic size (Sm, Eu, Ca) is key to a generally enhanced superconductivity irrespective of the details of the A-site composition and perhaps nor the substrate. With more systematic A-site compositional variation, which gives rise to different statistical variance in the distribution of A-site radii, cation effect on $T_c$ can be studied[47]. The large variation in ionic radius ($Sm^{3+}$ versus $Sr^{2+}$) may account for the lower $T_c$ in SSNO, likely an early indication to a 'size mismatch' scenario. The synthesis of an extended family of superconducting IL nickelates towards smaller lattice spacings offers an ideal opportunity to investigate such cation disorder effect in a nickelate system, despite that the lattice strain imposed by the substrate may vary across different IL compounds and intricately contribute to determining $T_c$. Furthermore, previous studies have yet failed to observe superconductivity in IL phases on substrates of smaller lattice constants, such as $LaAlO_3$, perhaps due to overall relatively large lattice mismatch between the IL nickelates and $LaAlO_3$. With a reducing lattice parameter of Sm-based compounds, a successful synthesis of a *c*-axis oriented IL phase on those substrates can be expected and warrants further investigation.

Additional interest lies in the possible salient features induced by magnetic $Eu^{2+}$ and/or $Sm^{3+}$, the latter of which possesses a weak magnetic moment, and their potential interaction with superconductivity. In particular, the half-filled 4*f* orbitals of $Eu^{2+}$ (total angular momentum $J$ = 7/2) produce highly localized magnetic moments (~ 7 μB) that participate in a Kondo-type spin-flip scattering with itinerant electrons within the $NiO_2$ layers. Upon external magnetic field, the alignment of these local moments may suppress the spin-disorder scattering, thereby reducing resistivity, which manifests as a negative magnetoresistance. In contrast, the weaker spin-charge interaction for $Sm^{3+}$ ($4f^5$, $J$ = 5/2) stems from the significantly smaller magnetic moment (~ 0.85 μB), hence the absence of the negative magnetoresistance. These unique configurations, bound to the local moments of 4*f* electrons, provide a versatile playground for the study of high-$T_c$ superconductivity (itinerate electrons) in a Kondo-lattice (local electrons) setting. In addition, as observed in the $Nd_{1-x}Eu_xNiO_2$ system[26], approximately 40 % of the Eu ions remain in the +3 state: Whether these $Eu^{2+}$/$Eu^{3+}$ ions form an ordered phase and/or exhibit a non-trivial magnetic ground state present interesting questions to studies like resonant inelastic X-ray scattering[26,61,64-67].

In summary, we have prepared superconducting $Sm_{1-x}Sr_xNiO_2$ thin films, a new single-dopant rare-earth IL nickelate using pulsed laser deposition. We have also synthesized an extended family of Sm-based co-hole-doped superconducting IL nickelate thin films with different dopants and enhanced superconductivity on the $(LaAlO_3)_{0.3}(Sr_2TaAlO_6)_{0.7}$ substrate. All samples show uniform coherent lattice structures with thickness of ~ 9 nm and were made from the fully metallic perovskite precursor phase despite a stringent growth parameter space. Particularly, due to an overall smaller lattice constant and a better cationic size-match, our doped Sm-based superconducting IL nickelates show an enhanced $T_c$ as high as 32.5 K. Rotational high-magnetic-field measurements consistently reveal a hybrid mixture of two- and three- dimensional superconductivity across all samples, which aligns with an enhanced interlayer coupling between the rare-earth 5*d* and Ni 3*d* orbitals, as further supported by the resonant X-ray scattering data. In comparison with the La-, Pr- and Nd-based IL compounds, our findings reveal a fundamental correlation where $T_c$ increases as the *c*-axis parameter decreases due to smaller ionic radii. These results advance our understanding of the superconducting mechanism in this system and encourage clear design principles for discovering additional IL nickelate superconductors with enhanced $T_c$.

## Methods

### Thin-film Growth

The polycrystalline $Sm_{1-x-y-z}Ca_xSr_yEu_zNiO_\delta$ ceramic targets were prepared by pelletising a mixture of $Sm_2O_3$, $Eu_2O_3$, $SrCO_3$, $CaCO_3$, and NiO powders, followed by decarbonisation step at 1250 °C for 12 hours. After that, the targets were ground, re-pelletised and then sintered twice at 1,300 °C and 1270 °C for 12 hours each. A slightly lower temperature for the final sintering was used to avoid a large volume change of the target due to thermal cycles. Using the targets, thin films with a thickness of ~ 10 nm were deposited on LSAT (001) substrates using a KrF excimer laser ($\lambda$ = 248 nm) in a pulsed laser deposition system. The growth temperature is 600 °C measured by a thermocouple. The oxygen pressure was maintained at 100 mTorr during the growth. The laser fluence was 2.6 $J/cm^2$ and the repetition frequency is 0.2 – 0.3 Hz on growth Sr-doped Sm-based nickelate films and 3 Hz repetition frequency for other films. It is noted that this frequency was adopted after a careful study on the frequency dependence and is more suitable for the phase formation of the precursor nickelates. High repetition frequency will lead to a diminution in the peak intensity of the (001) peak and induce a shift of the (002) peak position towards a lower angular value. A low laser fluence of 0.6 $J/cm^2$ was used to grow the epitaxial $SrTiO_3$ capping layer of ~ 2 nm.

### Topochemical Reduction

The reduction to the IL phase using $CaH_2$ powders was conducted in a vacuum reduction chamber. The resistance of the sample was monitored in a two-probe configuration in real-time. When the resistance reaches the minimum, the reduction is regarded as complete. The reduction temperature measured by a thermocouple is ~ 270 °C – 310 °C and the total annealing time is

~ 0.5 - 1.5 hours. Detailed information is provided in the Supplementary Information.

**X-ray Diffraction and RIXS Characterisations**

The X-ray diffraction $\theta - 2\theta$ symmetric scans of the films were obtained by a Rigaku SmartLab (9 kW) high-resolution X-ray diffractometer with the wavelength of the X-ray being 0.154 nm. The Ni $L$-edge RIXS measurements were conducted at the 41A beamline of the Taiwan Photon Source (TPS)[68]. All RIXS spectra were collected at $T = 30$ K, with a grazing-incidence angle of 10° and a scattering angle of 150°. The energy resolution, determined from the elastic scattering of amorphous carbon tape, was ~ 68 meV.

**STEM Sample Preparation and Characterisation**

Cross-sectional STEM specimens were prepared using a focused ion beam (FIB) technique with the Helios G4 system. To protect the sample surface during ion beam etching, a 2 $\mu$m-thick carbon layer was deposited beforehand. The preparation process involved a series of milling and lift-out steps, after which the lamellae were carefully extracted and mounted onto Cu grids for further characterization. To minimize surface damage, a final milling step was performed at 2 keV. HAADF images were acquired using an aberration-corrected FEI Titan Themis G2 operating at 300 kV. For STEM imaging, the convergence semi-angle was set to 30 mrad, while the collection semi-angle for HAADF imaging ranged from 50 to 200 mrad.

**In-house Transport Measurements**

Wire connections for electrical transport measurements using the standard four-probe method were made by aluminum ultrasonic wire-bonded contacts. For measuring perovskite precursor phase, gold wires were bonded to the sample with silver paste to avoid work function mismatch. Resistivity and Hall-effect measurements were performed at temperatures down to 2 K and under magnetic fields up to 14 T using a Physical Properties Measurement System from Quantum Design Inc. Two-coil mutual inductance experiments were conducted with the driving and pickup coils aligned vertically above and below the thin film samples. Polar angle dependent measurement is done on a rotation rod with Keithley 6220&2182A system.

**High-magnetic-field Measurements**

High-field magnetoresistance measurements were performed at the High Magnetic Field Laboratory, Chinese Academy of Sciences, Hefei, China, using the Steady High Magnetic Field Facilities (WM5) [69] with a maximum field of 35 T and a base temperature of 1.5 K. Resistance was measured using SR830 lock-in amplifiers with a built-in a.c. current source. $H_{c2,50\%}$ is defined as the point at which the magnetoresistance $\rho(H)$ curves intersect with an adjusted linear fit. Similar to the definition for $\rho(T)$ in Figure S5, this adjusted fit is attained by incorporating 50% of both the slope and the intercept of the linear fit to the normal-state data at $\theta = 0°$.

## Reference

1. Li, D. *et al.* Superconductivity in an infinite-layer nickelate. *Nature* **572**, 624–627 (2019).
2. Wang, B. Y., Lee, K. & Goodge, B. H. Experimental progress in superconducting nickelates. *Annu. Rev. Condens. Matter Phys* **15**, 305–324 (2024).
3. Gu, Q. & Wen, H.-H. Superconductivity in nickel-based 112 systems. *Innovation* **3**, (2022).
4. Keimer, B., Kivelson, S. A., Norman, M. R., Uchida, S. & Zaanen, J. From quantum matter to high-temperature superconductivity in copper oxides. *Nature* **518**, 179–186 (2015).
5. Hayward, M. A. Topochemical reactions of layered transition-metal oxides. *Semicond. Sci. Technol.* **29**, 064010 (2014).
6. Hepting, M. *et al.* Electronic structure of the parent compound of superconducting infinite-layer nickelates. *Nat. Mater.* **19**, 381–385 (2020).
7. Goodge, B. H. *et al.* Doping evolution of the Mott–Hubbard landscape in infinite-layer nickelates. *Proc. Natl. Acad. Sci.* **118**, e2007683118 (2021).
8. Lee, K.-W. Infinite-layer $LaNiO_2$ $Ni^{1+}$ is not $Cu^{2+}$. *Phys. Rev. B* **70**, 165109 (2004).
9. Sun, W. *et al.* Electronic structure of superconducting infinite-layer lanthanum nickelates. *Sci. Adv.* **11**, eadr5116 (2025).
10. Ding, X. *et al.* Cuprate-like electronic structures in infinite-layer nickelates with substantial hole dopings. *Nat. Sci. Rev.* **11**, nwae194 (2024).
11. Dong, Z. *et al.* Topochemical synthesis and electronic structure of high-crystallinity infinite-layer nickelates on an orthorhombic substrate. *Nano Lett.* **25**, 1233–1241 (2025).
12. Zhang, G.-M., Yang, Y. & Zhang, F.-C. Self-doped Mott insulator for parent compounds of nickelate superconductors. *Phys. Rev. B* **101**, 020501 (2020).
13. Jiang, P., Si, L., Liao, Z. & Zhong, Z. Electronic structure of rare-earth infinite-layer $RNiO_2$ (R=La,Nd). *Phys. Rev. B* **100**, 201106 (2019).
14. Bandyopadhyay, S., Adhikary, P., Das, T., Dasgupta, I. & Saha-Dasgupta, T. Superconductivity in infinite-layer nickelates: Role of f orbitals. *Phys. Rev. B* **102**, 220502 (2020).
15. Wang, B. Y. *et al.* Effects of rare-earth magnetism on the superconducting upper critical field in infinite-layer nickelates. *Sci. Adv.* **9**, eadf6655 (2023).
16. Lee, K. *et al.* Aspects of the synthesis of thin film superconducting infinite-layer nickelates. *APL Mater.* **8**, 041107 (2020).
17. Osada, M. *et al.* Nickelate superconductivity without rare-earth magnetism: $(La,Sr)NiO_2$. *Adv. Mater.* **33**, 2104083 (2021).
18. Wei, W. *et al.* Large upper critical fields and dimensionality crossover of superconductivity in the infinite-layer nickelate $La_{0.8}Sr_{0.2}NiO_2$. *Phys. Rev. B* **107**, L220503 (2023).
19. Zeng, S. *et al.* Superconductivity in infinite-layer nickelate $La_{1-x}Ca_xNiO_2$ thin films. *Sci. Adv.* **8**, eabl9927 (2022).
20. Osada, M., Wang, B. Y., Lee, K., Li, D. & Hwang, H. Y. Phase diagram of infinite layer praseodymium nickelate $Pr_{1-x}Sr_xNiO_2$ thin films. *Phys. Rev. Mater.* **4**, 121801 (2020).

21. Wang, N. N. *et al.* Pressure-induced monotonic enhancement of $T_c$ to over 30 K in superconducting $Pr_{0.82}Sr_{0.18}NiO_2$ thin films. *Nat. Commun.* **13**, 4367 (2022).
22. Li, D. *et al.* Superconducting dome in $Nd_{1-x}Sr_xNiO_2$ infinite layer films. *Phys. Rev. Lett.* **125**, 027001 (2020).
23. Zeng, S. *et al.* Phase diagram and superconducting dome of infinite-layer $Nd_{1-x}Sr_xNiO_2$ thin films. *Phys. Rev. Lett.* **125**, 147003 (2020).
24. Lee, K. *et al.* Linear-in-temperature resistivity for optimally superconducting (Nd,Sr)$NiO_2$. *Nature* **619**, 288–292 (2023).
25. Lee, Y. *et al.* Synthesis of superconducting freestanding infinite-layer nickelate heterostructures on the millimetre scale. *Nat. Synth.*, doi:10.1038/s44160-024-00714-2.
26. Wei, W., Vu, D., Zhang, Z., Walker, F. J. & Ahn, C. H. Superconducting $Nd_{1-x}Eu_xNiO_2$ thin films using in situ synthesis. *Sci. Adv.* **9**, eadh3327 (2023).
27. Lechermann, F. Multiorbital processes rule the $Nd_{1-x}Sr_xNiO_2$ normal state. *Phys. Rev. X* **10**, 041002 (2020).
28. Wang, Z., Zhang, G.-M., Yang, Y. & Zhang, F.-C. Distinct pairing symmetries of superconductivity in infinite-layer nickelates. *Phys. Rev. B* **102**, 220501 (2020).
29. Chen, X. H. *et al.* Superconductivity at 43 K in $SmFeAsO_{1-x}F_x$. *Nature* **453**, 761–762 (2008).
30. Chen, G. F. *et al.* Superconductivity at 41 K and its competition with spin-density-wave instability in layered $CeO_{1-x}F_xFeAs$. *Phys. Rev. Lett.* **100**, 247002 (2008).
31. Torrance, J. B., Lacorre, P., Nazzal, A. I., Ansaldo, E. J. & Niedermayer, Ch. Systematic study of insulator-metal transitions in perovskites $RNiO_3$ (R=Pr,Nd,Sm,Eu) due to closing of charge-transfer gap. *Phys. Rev. B* **45**, 8209–8212 (1992).
32. Catalano, S. *et al.* Rare-earth nickelates $RNiO_3$: thin films and heterostructures. *Rep. Prog. Phys.* **81**, 046501 (2018).
33. Dagotto, E. Correlated electrons in high-temperature superconductors. *Rev. Mod. Phys.* **66**, 763–840 (1994).
34. Dagotto, E. Complexity in strongly correlated electronic systems. *Science* **309**, 257–262 (2005).
35. Taillefer, L. Scattering and pairing in cuprate superconductors. *Annu. Rev. Condens. Matter Phys* **1**, 51–70 (2010).
36. Fernandes, R. M. *et al.* Iron pnictides and chalcogenides: a new paradigm for superconductivity. *Nature* **601**, 35–44 (2022).
37. Zeng, Z., Guenzburger, D., Ellis, D. E. & Baggio-Saitovitch, E. M. Effect of magnetism on superconductivity in rare-earth compounds $RENi_2B_2C$. *Physica C Supercond.* **271**, 23–31 (1996).
38. Pan, G. A. *et al.* Superconductivity in a quintuple-layer square-planar nickelate. *Nat. Mater.* **21**, 160–164 (2022).
39. Sun, H. *et al.* Signatures of superconductivity near 80 K in a nickelate under high pressure. *Nature* **621**, 493–498 (2023).
40. Zhu, Y. *et al.* Superconductivity in pressurized trilayer $La_4Ni_3O_{10-\delta}$ single crystals. *Nature* **631**, 531–536 (2024).

41. Ko, E. K. *et al.* Signatures of ambient pressure superconductivity in thin film $La_3Ni_2O_7$. *Nature* **638**, 935–940 (2025).
42. Zhou, G. *et al.* Ambient-pressure superconductivity onset above 40 K in $(La,Pr)_3Ni_2O_7$ films. *Nature* **640**, 641–646 (2025).
43. Gao, Q. *et al.* Magnetic excitations in strained infinite-layer nickelate $PrNiO_2$ films. *Nat. Commun.* **15**, 5576 (2024).
44. Phillips, J. M. Substrate selection for high-temperature superconducting thin films. *J. Appl. Phys.* **79**, 1829–1848 (1996).
45. Sun, W. *et al.* Evidence for anisotropic superconductivity beyond pauli limit in infinite-layer Lanthanum nickelates. *Adv. Mater.* **35**, 2303400 (2023).
46. Chow, S. L. E., Luo, Z. & Ariando, Bulk superconductivity near 40 K in hole-doped $SmNiO_2$ at ambient pressure. *Nature*, doi: 10.1038/s41586-025-08893-4 (2025).
47. Attfield, J. P., Kharlanov, A. L. & McAllister, J. A. Cation effects in doped $La_2CuO_4$ superconductors. *Nature* **394**, 157–159 (1998).
48. Fujita, K., Noda, T., Kojima, K. M., Eisaki, H. & Uchida, S. Effect of disorder outside the $CuO_2$ planes on $T_c$ of copper oxide superconductors. *Phys. Rev. Lett.* **95**, 097006 (2005).
49. Tranquada, J. M., Sternlieb, B. J., Axe, J. D., Nakamura, Y. & Uchida, S. Evidence for stripe correlations of spins and holes in copper oxide superconductors. *Nature* **375**, 561–563 (1995).
50. Fujita, M., Goka, H., Yamada, K., Tranquada, J. M. & Regnault, L. P. Stripe order, depinning, and fluctuations in $La_{1.875}Ba_{0.125}CuO_4$ and $La_{1.875}Ba_{0.075}Sr_{0.05}CuO_4$. *Phys. Rev. B* **70**, 104517 (2004).
51. García-Muñoz, J. L., Suaaidi, M., Martínez-Lope, M. J. & Alonso, J. A. Influence of carrier injection on the metal-insulator transition in electron- and hole-doped $R_{1-x}A_xNiO_3$ perovskites. *Phys. Rev. B* **52**, 13563–13569 (1995).
52. Cheong, S.-W., Hwang, H. Y., Batlogg, B., Cooper, A. S. & Canfield, P. C. Electron-hole doping of the metal-insulator transition compound $RENiO_3$. *Physica B* **194–196**, 1087–1088 (1994).
53. Stankiewicz, J., Bartolomé, J. & Fruchart, D. Spin disorder scattering in magnetic metallic alloys. *Phys. Rev. Lett.* **89**, 106602 (2002).
54. Shannon, R. D. Revised effective ionic radii and systematic studies of interatomic distances in halides and chalcogenides. *Acta. Cryst. A* **32**, 751–767 (1976).
55. He, X., Gozar, A., Sundling, R. & Božović, I. High-precision measurement of magnetic penetration depth in superconducting films. *Review of Scientific Instruments* **87**, 113903 (2016).
56. Wang, B. Y. *et al.* Isotropic Pauli-limited superconductivity in the infinite-layer nickelate $Nd_{0.775}Sr_{0.225}NiO_2$. *Nat. Phys.* **17**, 473–477 (2021).
57. Xiang, Y. *et al.* Physical properties revealed by transport measurements for superconducting $Nd_{0.8}Sr_{0.2}NiO_2$ thin films. *Chin. Phys. Lett* **38**, 047401–047401 (2021).
58. Palstra, T. T. M., Batlogg, B., Schneemeyer, L. F. & Waszczak, J. V. Thermally Activated Dissipation in $B_{2.2}Sr_2Ca_{0.8}Cu_2O_{8+\delta}$. *Phys. Rev. Lett.* **61**, 1662–1665 (1988).
59. Chow, L. E. *et al.* Pauli-limit violation in lanthanide infinite-layer nickelate superconductors. arXiv [Preprint] doi:10.48550/arXiv.2204.12606.

60. Rossi, M. *et al.* Orbital and spin character of doped carriers in infinite-layer nickelates. *Phys. Rev. B* **104**, L220505 (2021).
61. Rossi, M. *et al.* Universal orbital and magnetic structures in infinite-layer nickelates. *Phys. Rev. B* **109**, 024512 (2024).
62. Krockenberger, Y. *et al.* Superconductivity phase diagrams for the electron-doped cuprates $R_{2-x}Ce_xCuO_4$ (R=La, Pr, Nd, Sm, and Eu). *Phys. Rev. B* **77**, 060505 (2008).
63. Been, E. *et al.* Electronic structure trends across the rare-earth series in superconducting infinite-layer nickelates. *Phys. Rev. X* **11**, 011050 (2021).
64. Lu, H. *et al.* Magnetic excitations in infinite-layer nickelates. *Science* **373**, 213–216 (2021).
65. Rossi, M. *et al.* A broken translational symmetry state in an infinite-layer nickelate. *Nat. Phys.* **18**, 869–873 (2022).
66. Hayashida, S. *et al.* Investigation of spin excitations and charge order in bulk crystals of the infinite-layer nickelate $LaNiO_2$. *Phys. Rev. B* **109**, 235106 (2024).
67. Parzyck, C. T. *et al.* Absence of $3a_0$ charge density wave order in the infinite-layer nickelate $NdNiO_2$. *Nat. Mater.* **23**, 486–491 (2024).
68. Singh, A. *et al.* Development of the Soft X-ray AGM–AGS RIXS beamline at the Taiwan Photon Source. *J Synchrotron Radiat* **28**, 977–986 (2021).
69. Fang, Z. et al. Design of a 42 T Resistive Magnet at the CHMFL, IEEE *Trans. Appl. Superconduct*., 34, 1-4 (2024).

## Acknowledgements

We thank Ariando and Lin Er Chow for discussions. We acknowledge the funding support from the National Natural Science Foundation of China (12174325) and a Guangdong Basic and Applied Basic Research Grant (2023A1515011352). The research was supported by research grants from the Research Grants Council (RGC) of the Hong Kong Special Administrative Region, China, under Early Career Scheme, General Research Fund and ANR-RGC Joint Researh Scheme (CityU 21301221, CityU 11309622, CityU 11300923, A-CityU102/23 and CUHK 24306223). Part of the work utilized the equipment support through a Collaborative Research Equipment Grant from RGC (C1018-22E). Part of this work was supported by the National Key R&D Program of China (2024YFA1408101 and 2022YFA1403101), the Natural Science Foundation of China (92265112, 12374455 and 52388201), the Guangdong Provincial Quantum Science Strategic Initiative (GDZX2401004, GDZX2201001 and GDZX2401012), the Shenzhen Science and Technology Program (KQTD20240729102026004), and the Shenzhen Municipal Funding Co-Construction Program Project (SZZX2301004 and SZZX2401001). The high-magnetic-field work was supported by the National Key R&D Program of China (2023YFA1607701) and National Natural Science Foundation of China (51627901). We thank the staff members of the SMA and WM5 System at the Steady High Magnetic Field Facility, Chinese Academy of Sciences, for providing technical support and assistance in data collection and analysis. The resonant X-ray scattering experiments were supported by National Science and Technology Council (NSTC), Taiwan (Grant No. 113-2112-M-213-016). P.G. acknowledges the support from the New Cornerstone Science Foundation through the XPLORER PRIZE. We acknowledge Electron Microscopy Laboratory of Peking University for the use of electron microscopes.

**Author contributions**

M.Y., H.W. and J.T. contributed equally to this work. M.Y., Z.C. and D.L. conceived the research project. M.Y. and J.T. grew the samples with assistance from W.X., Z.D., B.F., L.S. and Z.P. H.W. and X.W. performed the in-situ reduction experiments. H.W., X.W., M.Y. and G.Z. performed the XRD characterizations. H.W. conducted the mutual inductance measurements, H.W. and J.T. performed the transport measurements. J.L., R.M. and P.G. conducted the STEM experiments. A.W., Z.W., M.Y. and H.H. conducted the high field measurements with the help from W.M., C.X., L.P. and Q.L. Y.W. and Q.W. conducted the RIXS measurements with the help from J.O., H.-Y.H. and D.-J.H. H.H., Q.W., P.G., Z.C. and D.L. acquired funding support. M.Y., J.T., W.X. and D.L. wrote the manuscript with contribution from all authors.

**Competing interests**

The authors declare that they have no competing interests.

**Additional information**

Supplementary information: The online version contains supplementary materials available at (to be filled).

The authors declare that the data supporting the findings of this study are available within the paper, its supplementary information files, and from the corresponding authors on reasonable request.

Correspondence and requests for materials should be addressed to Zhuoyu Chen or Danfeng Li

## Figures

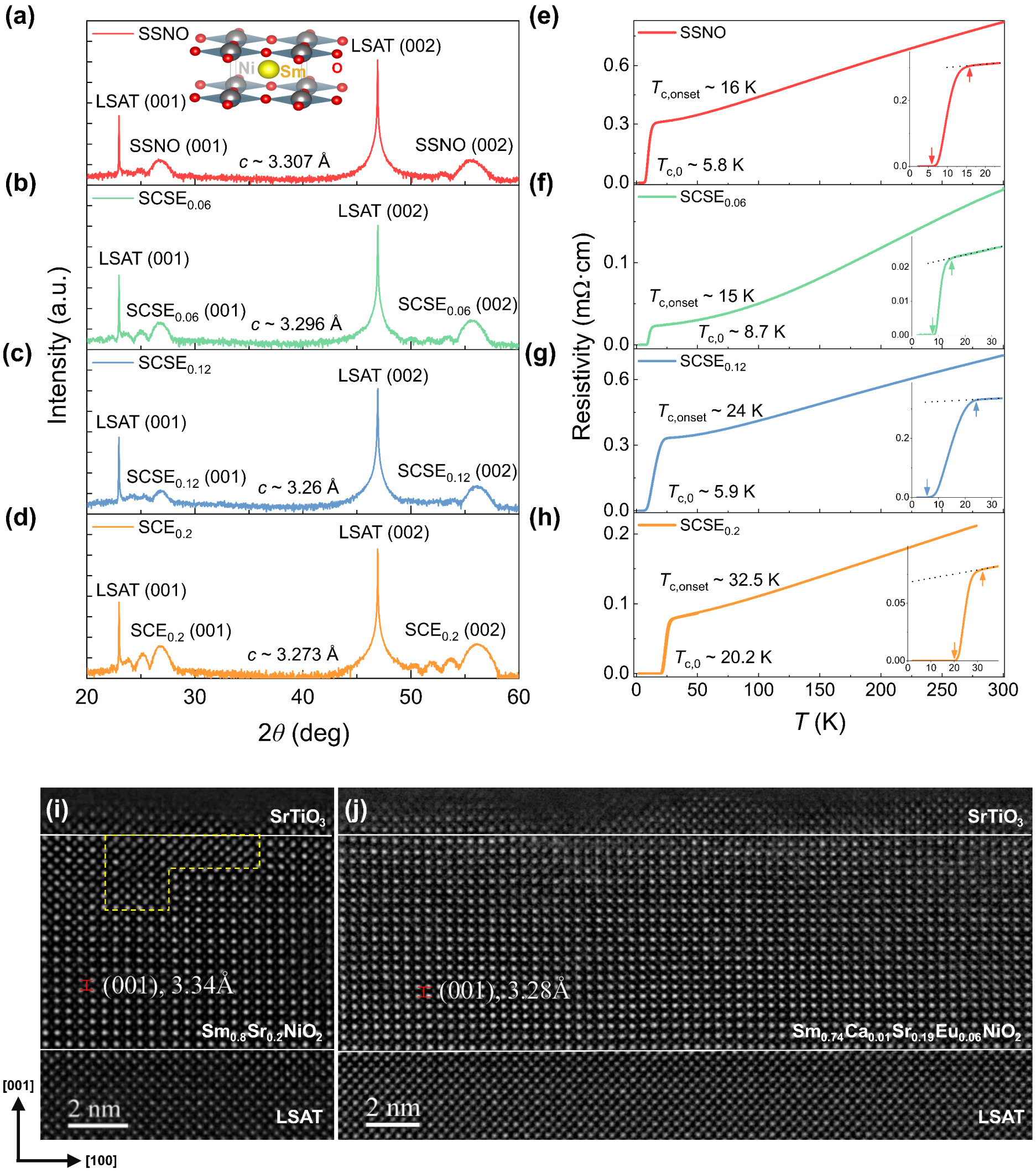

**Figure 1**. (**a**), (**b**), (**c**) and (**d**) are the X-ray diffraction (XRD) $\theta$–$2\theta$ symmetric patterns of $Sm_{0.8}Sr_{0.2}NiO_2$ (SSNO), $Sm_{0.74}Ca_{0.01}Sr_{0.19}Eu_{0.06}NiO_2$ ($SCSE_{0.06}$), $Sm_{0.79}Ca_{0.04}Sr_{0.05}Eu_{0.12}NiO_2$ ($SCSE_{0.12}$) $Sm_{0.75}Ca_{0.05}Eu_{0.2}NiO_2$ ($SCE_{0.2}$) thin films. Inset of (**a**) is a schematic diagram of the atomic structure. Resistivity curves $\rho(T)$ of SSNO, $SCSE_{0.06}$, $SCSE_{0.12}$ and $SCE_{0.2}$ samples are shown in (**e**), (**f**), (**g**) and (**h**). Insets of (**e**), (**f**), (**g**) and (**h**) show the zoom-in data around the superconducting transitions. Dashed lines are linear fits to the normal state $\rho(T)$ curves above the transitions. Here, $T_{c,onset}$ is defined as the point where the curve deviates from the linear fitting, $T_{c,0}$ is defined as the zero-resistance $T_c$. (**i**) and (**j**) are high-angle annular dark-field (HAADF) images of an SSNO thin film and an $SCSE_{0.06}$ thin film. The $c$-axis lattice constants are measured to be 3.34 Å for SSNO and 3.28 Å for $SCSE_{0.06}$. The area circulated by yellow dashed line indicates a Ruddlesden-Popper (RP) stacking fault.

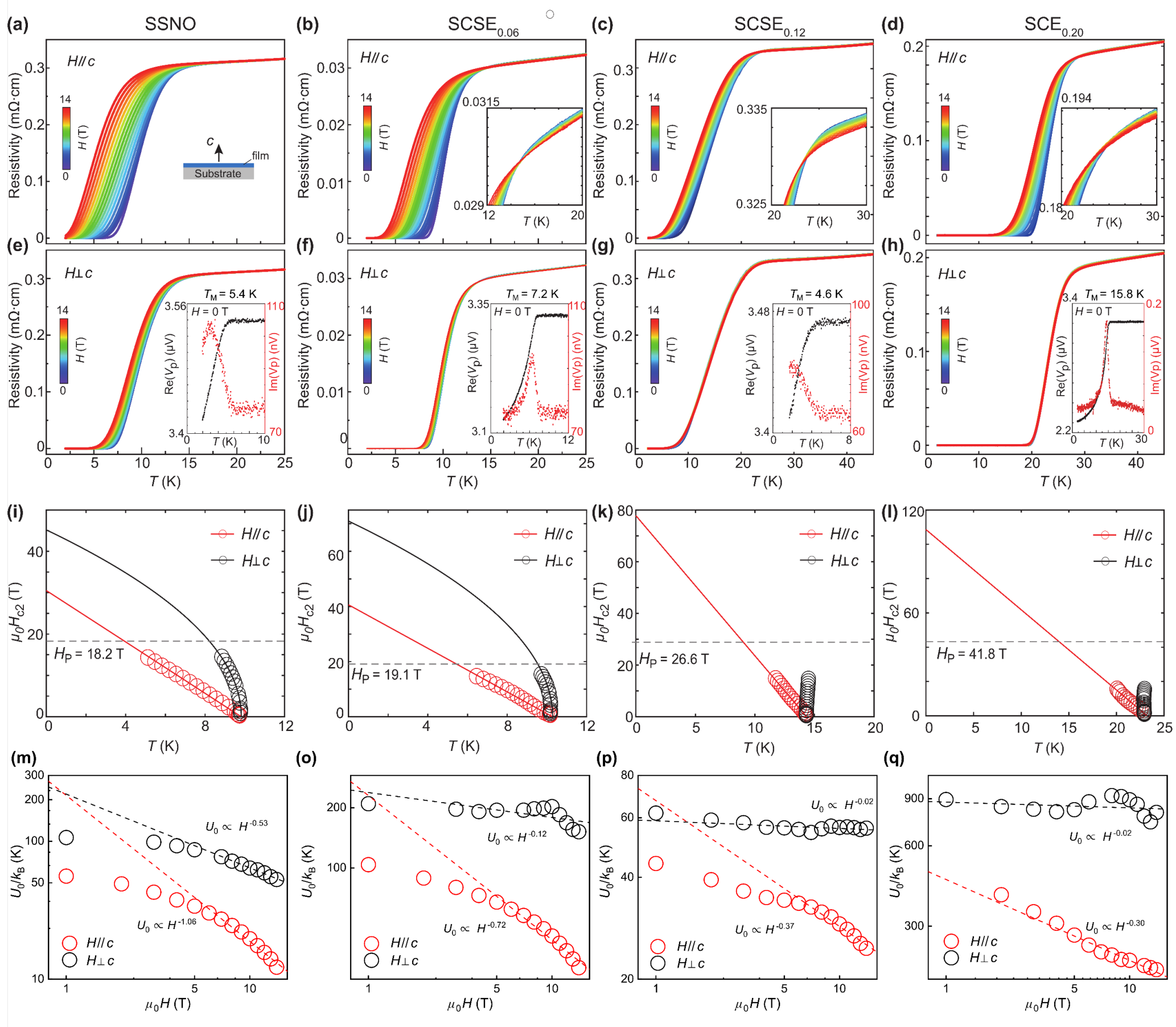

**Figure 2.** Magnetic-field responses of the superconducting SSNO, $SCSE_{0.06}$, $SCSE_{0.12}$ and $SCE_{0.2}$ thin films. Figure (**a**), (**b**), (**c**), (**d**) ((**e**), (**f**), (**g**), (**h**)) are $\rho(T)$ under varying magnetic field perpendicular (parallel) to the $NiO_2$ planes of SSNO, $SCSE_{0.06}$, $SCSE_{0.12}$ and $SCE_{0.2}$ films. Inset of (**a**) is a schematic of the sample cross-section. Insets of (**b**), (**c**) and (**d**) show the zoom-in data around the onset of the superconducting transitions: negative magnetoresistance can be seen above the transitions. Insets of (**e**), (**f**), (**g**), (**h**) show the mutual inductance results for each sample, where the superconducting diamagnetic response can be clearly observed. (**i**), (**j**), (**k**) and (**l**) show the variation of the upper critical field $\mu_0 H_{c2\perp}$ and $\mu_0 H_{c2//}$ (estimated against the $T_c$ definition described in the main text) of SSNO, $SCSE_{0.06}$, $SCSE_{0.12}$ and $SCE_{0.2}$ fitted with the Ginzburg-Landau (G-L) equations, where applicable. (**m**), (**n**), (**o**), and (**p**) show the vortex activation energy $U_0/k_B$ as a function of magnetic field $\mu_0 H$ for the SSNO, $SCSE_{0.06}$, $SCSE_{0.12}$ and $SCE_{0.2}$ thin films under magnetic fields applied parallel and perpendicular to the $NiO_2$ planes. The data are fitted with a power-law relationship $U_0 \sim H^{-\alpha}$, extracted from the Arrhenius plot of $\rho(T)$ using the thermally activated flux flow (TAFF) model.

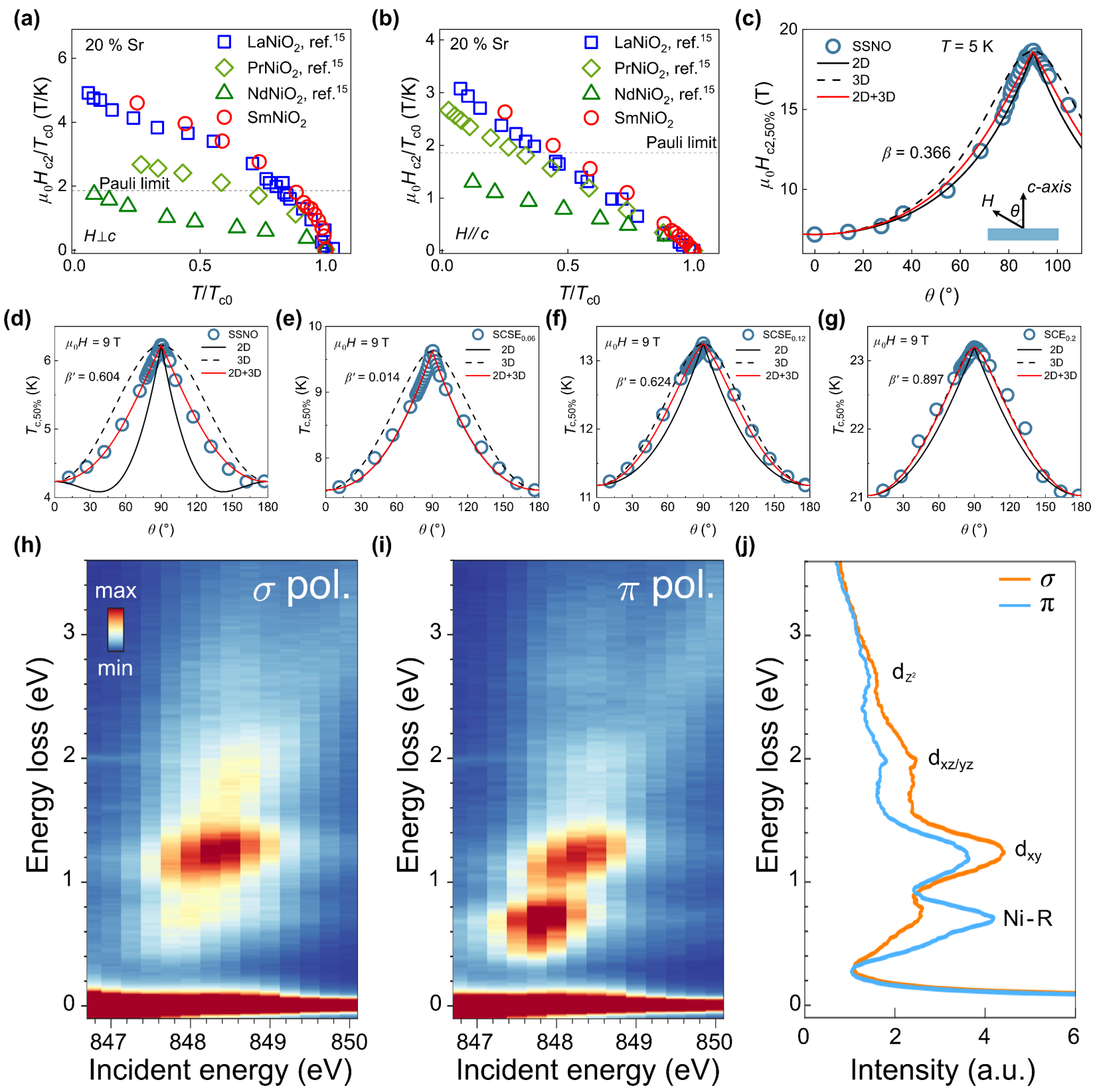

**Figure 3.** (**a**) and (**b**) are normalized $H_{c2}/T_{c0}$ against reduced temperature $T/T_{c0}$ of $RE_{0.8}Sr_{0.2}NiO_2$ (RE: La, Pr, Nd and Sm) for $H\perp c$ and $H//c$. The data for La-, Pr- and Nd- series are adapted from ref.[15]. The dashed lines indicate the Pauli limit of $H_{c2}$ = 1.86 $T_{c0}$. (**c**) shows the upper critical field $\mu_0 H_{c2,50\%}$ as a function of the polar rotation angle $\theta$ measured at 5 K. Here, the cyan line is a fit using the 2D Tinkham's model, while the dashed line is a fit using the anisotropic 3D GL model. The red line is a fit using both 2D and 3D models with the ratio of the 3D component, $\beta$. Inset of (**c**) is a schematic of the measurement geometry. (**d**), (**e**), (**f**) and (**g**) are $T_{c,50\%}$ as a function of $\theta$ measured at 9 T for the SSNO, $SCSE_{0.06}$, $SCSE_{0.12}$ and $SCE_{0.2}$ thin films, respectively. The red lines are fits to both 2D and 3D models with the ratio of the 3D component, $\beta'$. (**h**) and (**i**) are the resonant inelastic X-ray scattering (RIXS) energy loss maps as a function of incident energy for two light polarizations. (**j**) is the integrated intensity for both light polarizations.

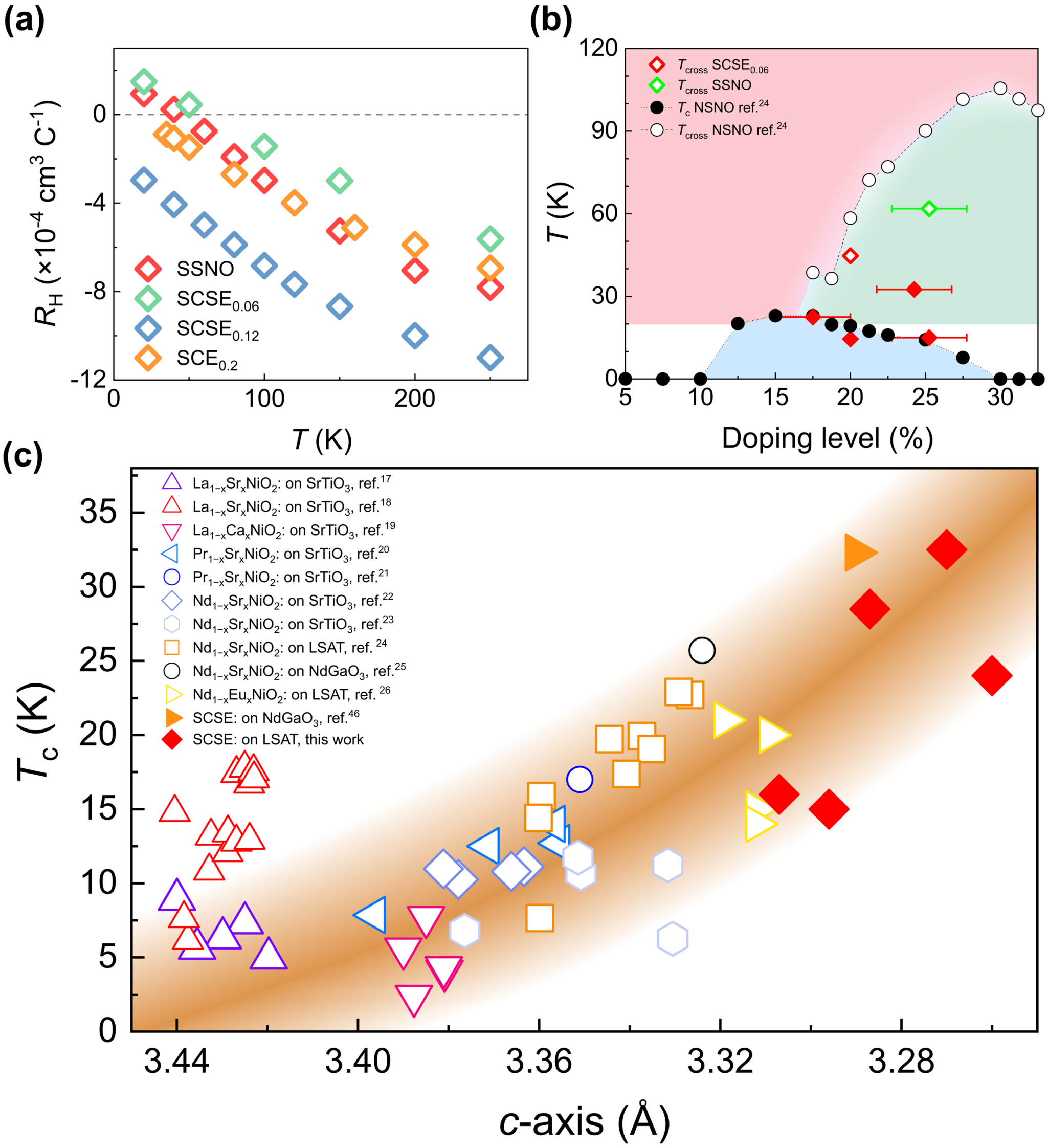

**Figure 4**. (a) Temperature dependence of the Hall coefficients for various samples. (b) $T_c$ and $T_{cross}$ versus the estimated hole doping level plotted with reference to the data extracted from ref.[24] for $Nd_{1-x}Sr_xNiO_2$ (NSNO). (c) The correlation of $T_c$ and the $c$-axis lattice constant for different IL systems. The highest $T_c$ was observed in $SCE_{0.2}$ with a $c$-axis of 3.273 Å. The red rhombus-shaped points in the figure are data from this work while other data are extracted from refs.[17–26,46]

**Supplementary Information *for***

# Enhanced Superconductivity and Mixed-dimensional Behaviour in Infinite-layer Samarium Nickelate Thin Films

Mingwei Yang[1,2,12], Heng Wang[3,12], Jiayin Tang[1,12], Junping Luo[4], Xianfeng Wu[3], Wenjing Xu[1], Aile Wang[5,6], Yuetong Wu[7], Ruilin Mao[4], Ze Wang[8], Zhicheng Pei[1], Guangdi Zhou[3], Zhengang Dong[1,2], Bohan Feng[1,2], Lingchi Shi[1], Wenjie Meng[8], Chuanying Xi[8], Li Pi[8], Qingyou Lu[5,6,8], Jun Okamoto[9], Hsiao-Yu Huang[9], Di-Jing Huang[9], Haoliang Huang[3,10], Qisi Wang[7], Peng Gao[4,11], Zhuoyu Chen[3,10*] and Danfeng Li[1,2*]

[1]*Department of Physics, City University of Hong Kong, Kowloon, Hong Kong SAR, China.*
[2]*Shenzhen Research Institute of City University of Hong Kong, Shenzhen 518057, China.*
[3]*Department of Physics, State Key Laboratory of Quantum Functional Materials, and Guangdong Basic Research Center of Excellence for Quantum Science, Southern University of Science and Technology, Shenzhen 518055, China.*
[4]*International Center for Quantum Materials, and Electron Microscopy Laboratory, School of Physics, Peking University, Beijing 100871, China.*
[5]*Hefei National Research Center for Physics Sciences at the Microscale, University of Science and Technology of China; Hefei 230026, China.*
[6]*Hefei National Laboratory, University of Science and Technology of China; Hefei 230026, China.*
[7]*Department of Physics, The Chinese University of Hong Kong, Shatin, Hong Kong SAR, China.*
[8]*Anhui Key Laboratory of Low-Energy Quantum Materials and Devices, High Magnetic Field Laboratory, HFIPS, Chinese Academy of Sciences, Hefei 230031, China.*
[9]*National Synchrotron Radiation Research Center, Hsinchu 30076, Taiwan.*
[10]*Quantum Science Center of Guangdong-Hong Kong-Macao Greater Bay Area, Shenzhen 518045, China.*
[11]*Collaborative Innovation Center of Quantum Matter, Beijing 100871, China.*
[12]*These authors contributed equally*: Mingwei Yang, Heng Wang, Jiayin Tang.
[*]*Corresponding author*: chenzhuoyu@sustech.edu.cn, danfeng.li@cityu.edu.hk.

## I. Precursor Perovskite Sm-based Nickelates.

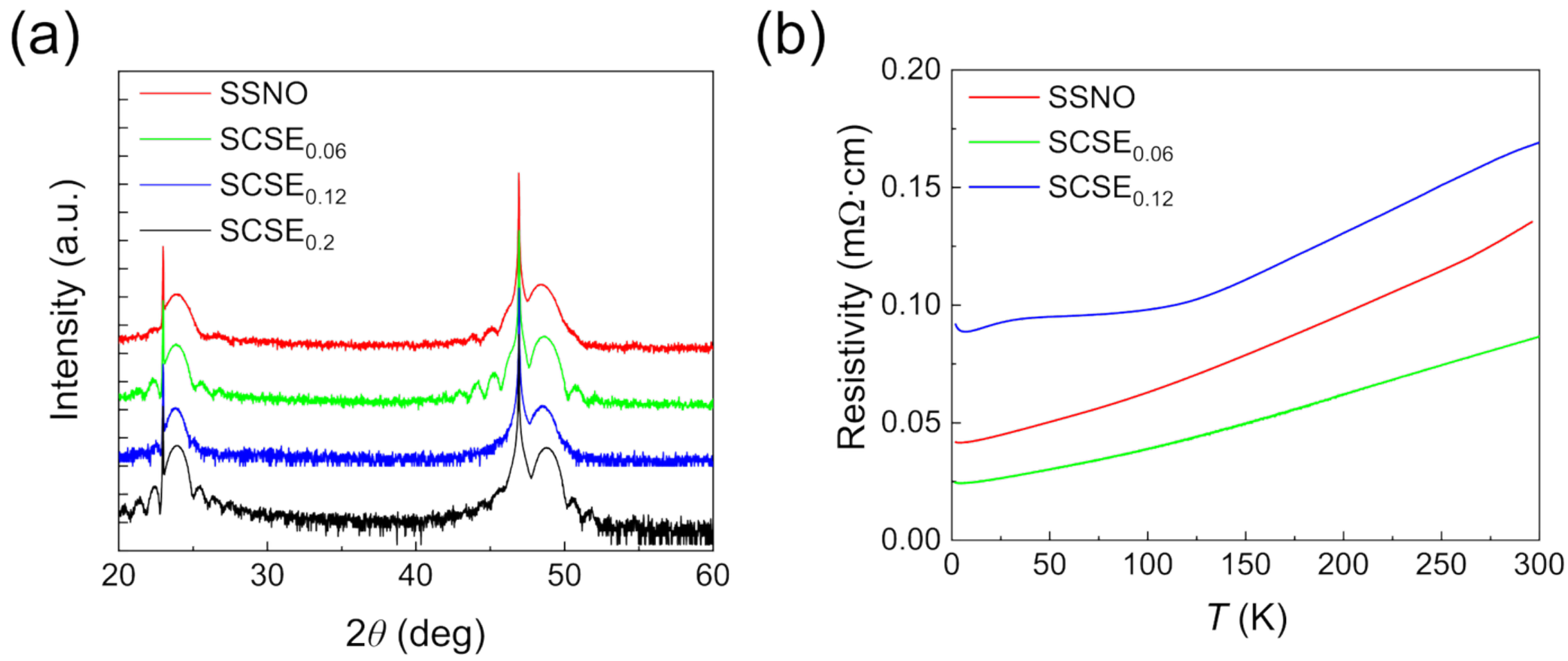

**Figure S1. The structural and transport characterizations of the as-grown samples.** (a) X-ray diffraction (XRD) $\theta$–$2\theta$ symmetric patterns of $Sm_{0.8}Sr_{0.2}NiO_3$, $Sm_{0.74}Ca_{0.01}Sr_{0.19}Eu_{0.06}NiO_3$, $Sm_{0.79}Ca_{0.04}Sr_{0.05}Eu_{0.12}NiO_3$ and $Sm_{0.75}Ca_{0.05}Eu_{0.2}NiO_3$ thin films. (b) The temperature-dependent resistivity for first three samples.

**Figure S1** shows the XRD patterns of the as-grown perovskite phases of the samples discussed in the main text and their corresponding $\rho(T)$ curves. The presence of prominent (001) film peaks with finite-size oscillations and their peak positions suggest good film crystallinity, which is further corroborated by the metallic behaviours of representative samples. For $Sm_{0.79}Ca_{0.04}Sr_{0.05}Eu_{0.12}NiO_3$, the local 'plateau' in resistivity at ~ 60 K may correspond to an antiferromagnetic ordering temperature of Sm, while the local resistivity minimum at ~ 8 K correlates with a spin-disorder-related metal-insulator transition induced by a partial substitution of Sm with Eu[1]. These transitions are more revealing on a derivative resistivity curve (**Figure S2**).

## II. Analysis of Resistivity Curve of $Sm_{0.79}Ca_{0.04}Sr_{0.05}Eu_{0.12}NiO_3$

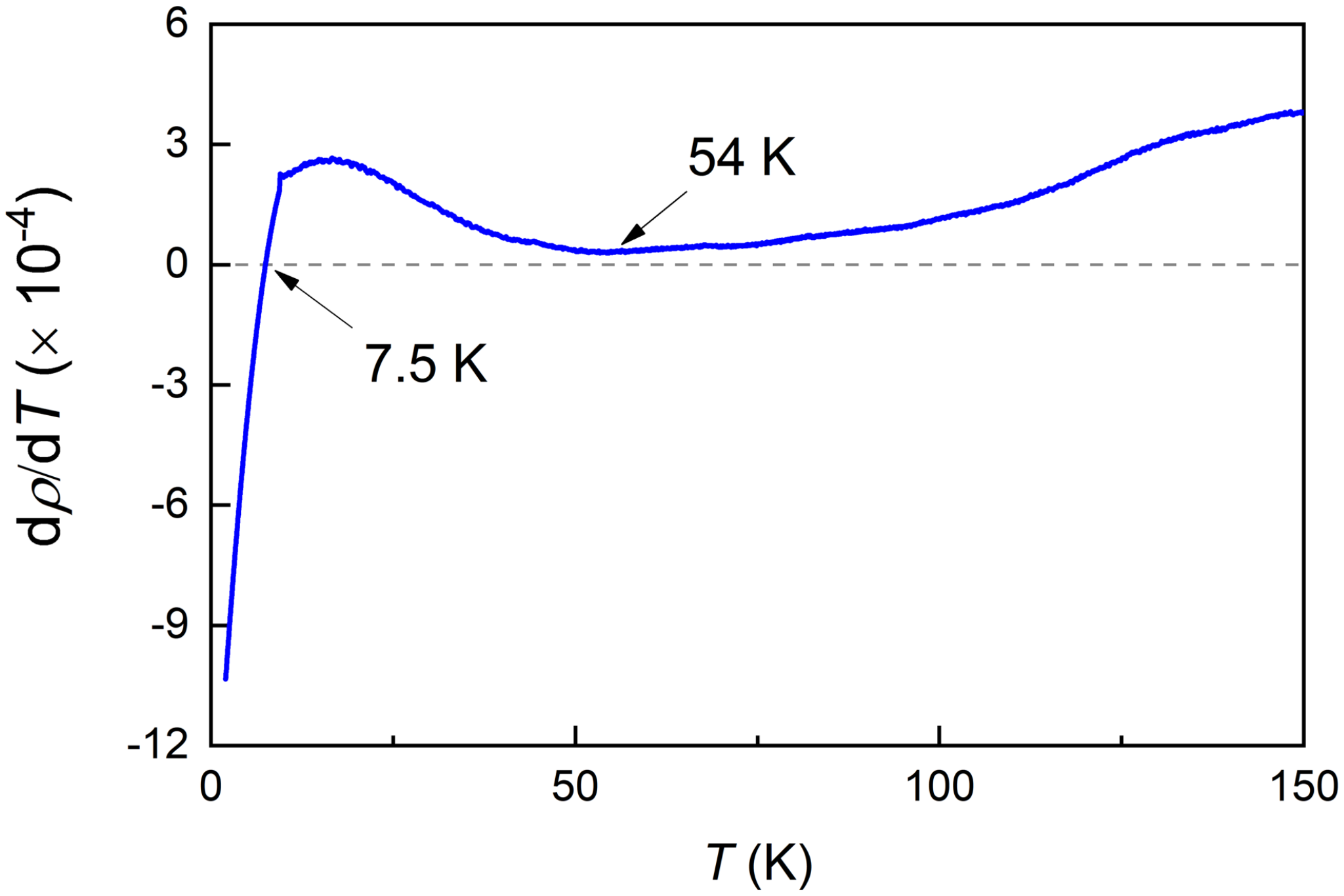

**Figure S2. First derivative of the temperature-dependent resistivity data of the $Sm_{0.79}Ca_{0.04}Sr_{0.05}Eu_{0.12}NiO_3$ sample from Fig. 1B in the main text.** There is a local minimum at 54 K and the slope reaches zero at 7.5 K.

Figure S2 shows the first derivative of the $\rho(T)$ data for the $Sm_{0.79}Ca_{0.04}Sr_{0.05}Eu_{0.12}NiO_3$ sample. The sign change at 7.5 K may be attributed to the spin disorder effects of Sm and Eu[1]. The change of concavity and convexity at 54 K may be linked to a magnetic ordering transition of Sm.

## III. Samples with Different Ca Doping.

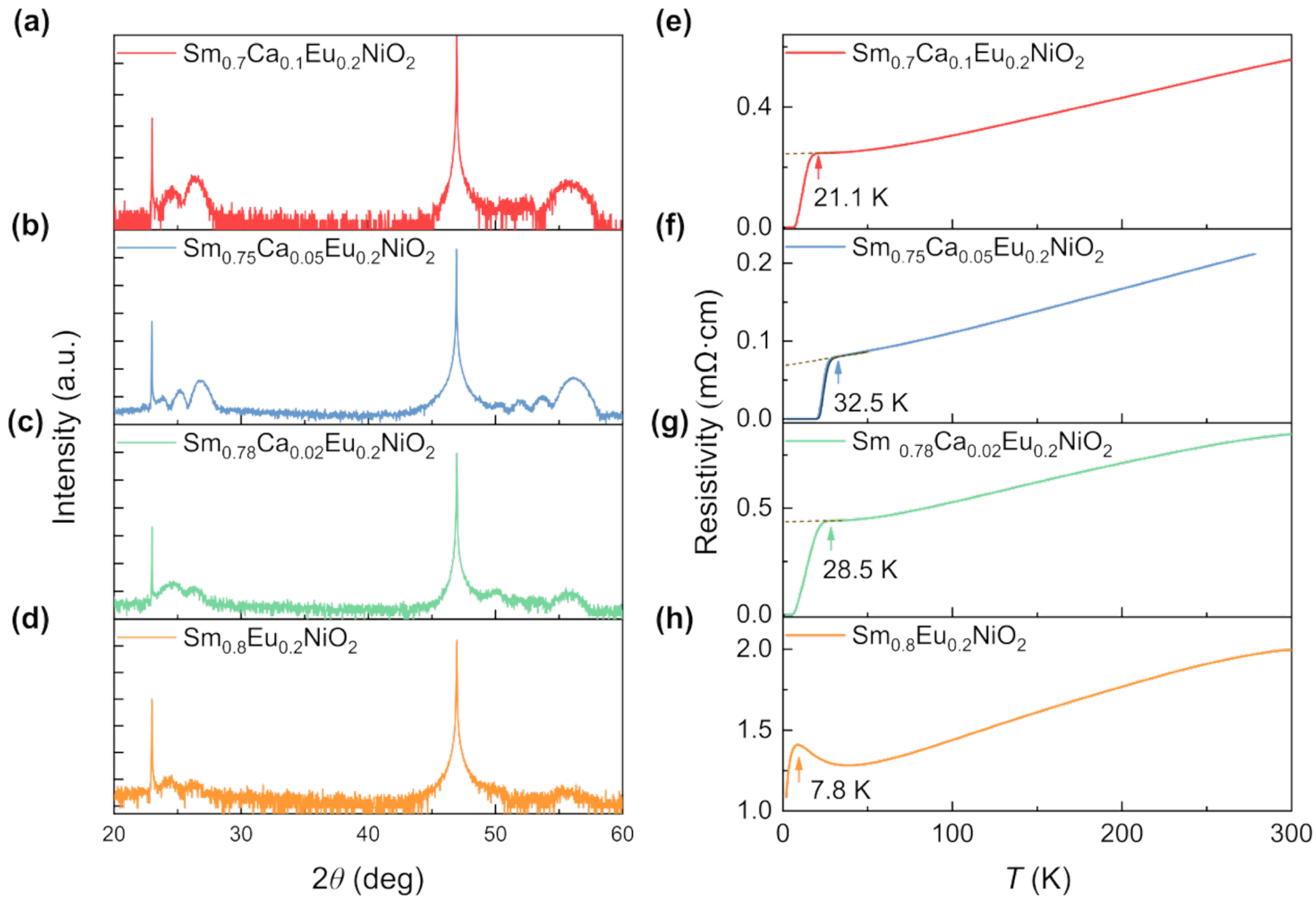

**Figure S3. Structural characterizations and electrical transport of samples with Different Ca Doping.** XRD $\theta$-$2\theta$ symmetric scans of $Sm_{0.7}Ca_{0.1}Eu_{0.2}NiO_2$, $Sm_{0.75}Ca_{0.05}Eu_{0.2}NiO_2$, $Sm_{0.78}Ca_{0.02}Eu_{0.2}NiO_2$ and $Sm_{0.8}Eu_{0.2}NiO_2$ films grown on LSAT (001) substrates (a-d). Superconductivity in $Sm_{0.7}Ca_{0.1}Eu_{0.2}NiO_2$, $Sm_{0.75}Ca_{0.05}Eu_{0.2}NiO_2$, $Sm_{0.78}Ca_{0.02}Eu_{0.2}NiO_2$ and $Sm_{0.8}Eu_{0.2}NiO_2$ (e-f) thin films. The data in (b) and (f) (blue curve) are the same dataset shown in Figure 1, d and h (brown curves).

We initially attempted to grow 20 % Eu-doped samarium nickelate samples, which exhibit high crystallinity but were challenging to reduce to an infinite-layer (112) phase. Replacing Sm with a trace amount of Ca significantly improved the reduction process, facilitating 112 phase formation. **Figure S3** presents the XRD data for $Sm_{0.8-x}Ca_xEu_{0.2}NiO_2$ ($x$ = 0, 0.02, 0.05 and 0.1) with various Ca substitution, showing the $c$-axis constants of 3.297 Å, 3.287 Å, 3.273 Å and 3.291 Å and corresponding superconducting transition temperatures ($T_c$) of 7.8 K, 28.5 K and 32.5 K and 21.1 K, respectively. Here, the data in Figures S3b and S3f for $Sm_{0.75}Ca_{0.05}Eu_{0.2}NiO_2$ have been shown in Figure 1. This is consistent with the main finding of our work that smaller lattice parameters lead to relatively higher $T_c$.

## IV. In-situ Reduction Process

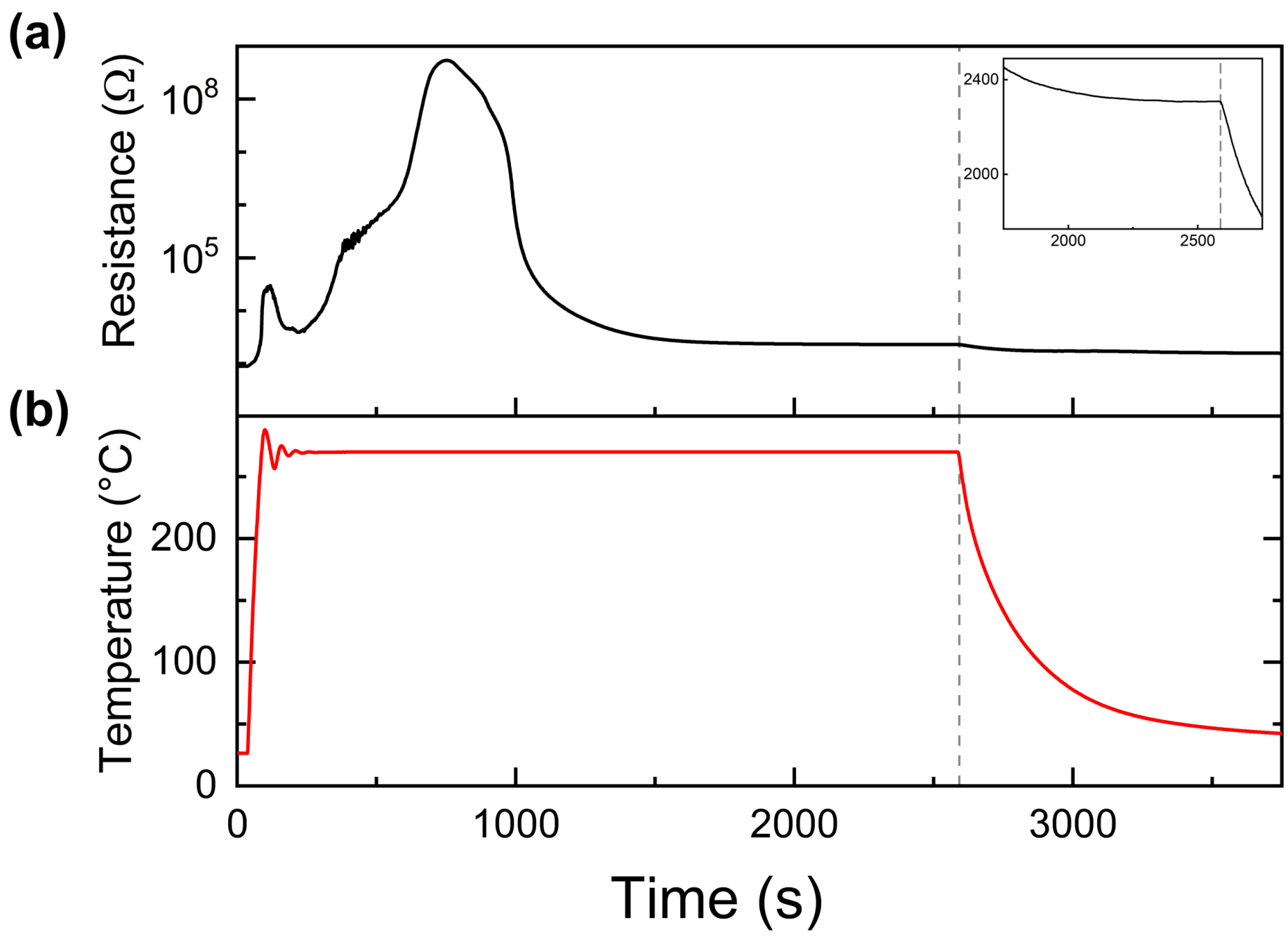

**Figure S4. Time-dependent reduction profiles of (a) resistance and (b) reduction temperature.** The inset of (a) shows the saturation point of resistance, indicating the completion of the reduction process.

The reduction process was performed in a vacuum reduction chamber using 1 g of $CaH_2$ powder. A Keithley 2450 source meter was used to monitor the sample with a two-probe configuration during the reduction. **Figure S4** presents the time-dependent resistance and temperature profiles during a typical reduction experiment. The complex temporal curve of resistance shown in (a) reflects different stages of the reduction reaction. The initial increase in resistance is caused by the early-stage oxygen de-intercalation, giving rise to an insulating phase. The subsequent decrease in resistance marks the onset of formation of the 112 structure. As shown in the inset of Figure S4a, the resistance reaches a saturation point at the reduction temperature, which is considered the endpoint of the reduction process, before the sample is cooled down.

## V. Definition of $\rho_{50\%}$

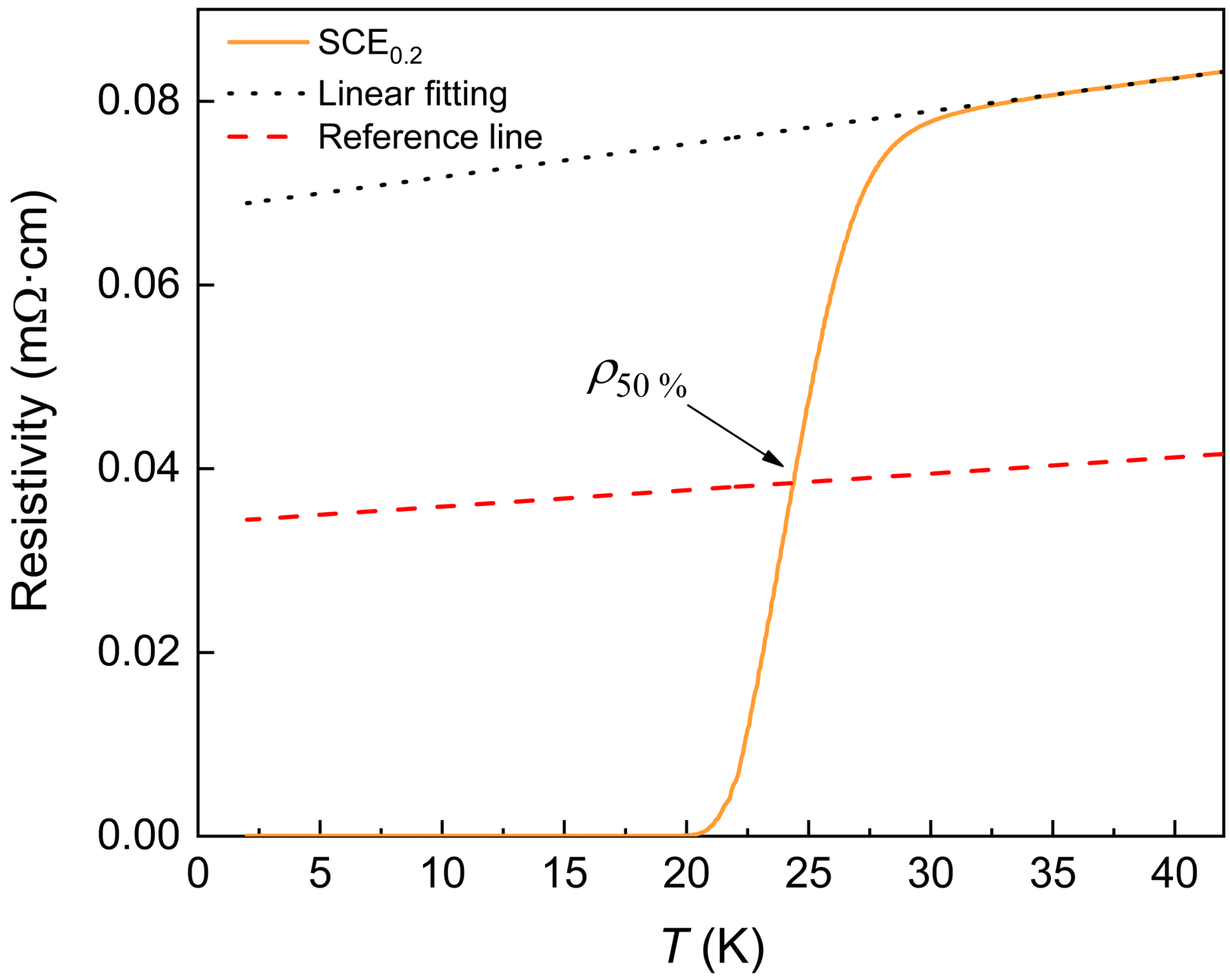

**Figure S5. Definition of $\rho_{50\%}$.** The brown curve represents the $\rho(T)$ data for $Sm_{0.75}Ca_{0.05}Eu_{0.2}NiO_2$ ($SCE_{0.2}$) as shown in the main text (Figure 1h). The black dotted line is the linear fitting to the data in the normal state. The red dashed line is derived from the linear fit, with 50 % of the slope and the intercept (to the resistivity axis). The intersection between the data and the red dashed line is defined as $\rho_{50\%}$.

Here, $\rho_{50\%}$ is defined as the intersection of the $\rho(T)$ curve and the adjusted linear fit, incorporating 50 % of both the slope and intercept (red dashed line) from the linear fitting to the normal-state data (black dotted line), as shown in **Figure S5**.

**VI. Activation Energy of the Vortex Motion**

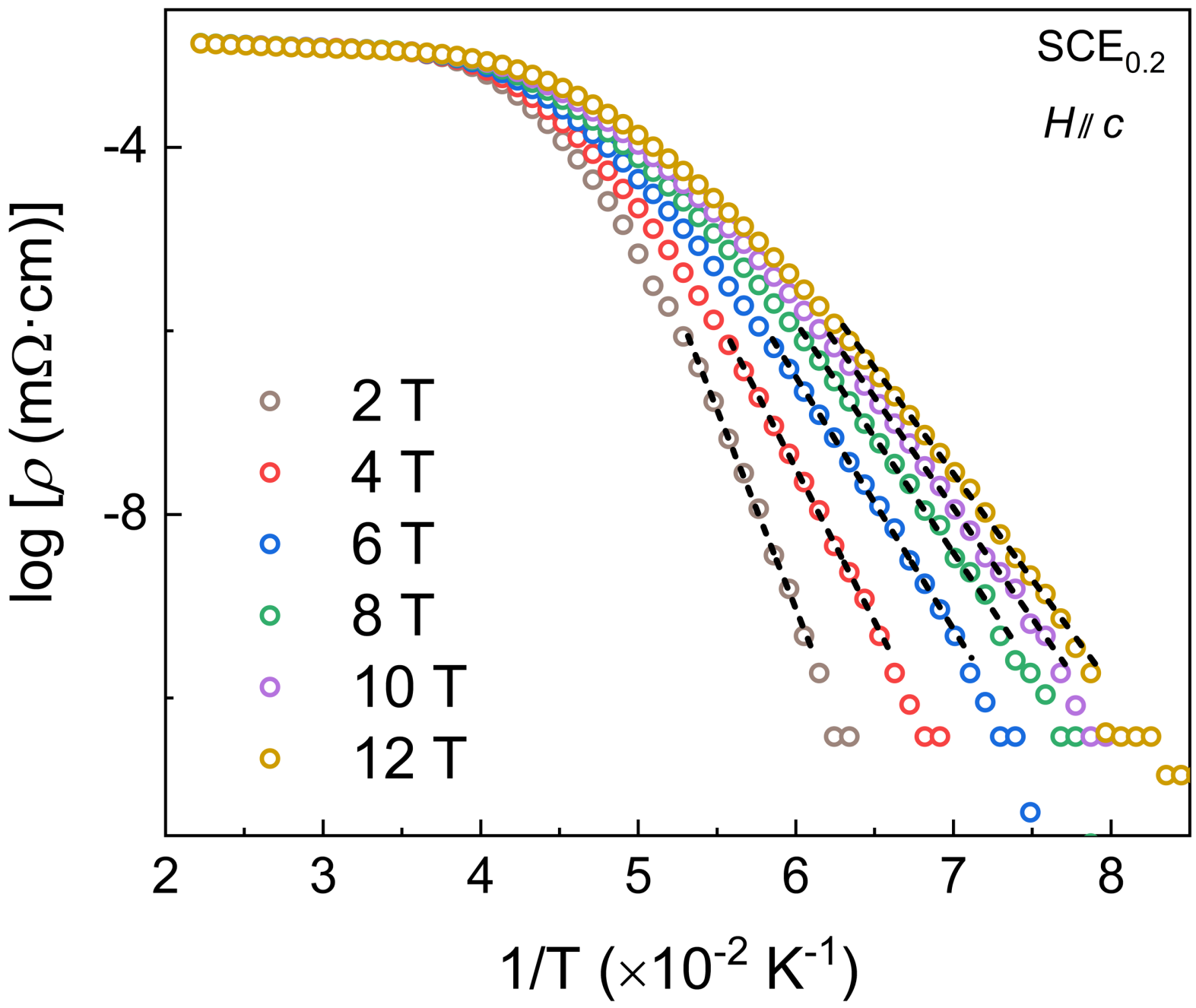

**Figure S6. Arrhenius fitting of the logarithmic resistivity versus inverse temperature and extraction of the thermal activation energy ($U_0$) for the vortex motion.** The representative sample is $SCE_{0.2}$.

The broadening of the superconducting transitions indicates a thermally activated behaviour of the superconducting vortices. Using a thermally-activated flux flow (TAFF) model[2], the field-dependent vortex activation energy, $U_0$ ($H$) can be extracted and plotted (**Figure S6**, and $U_0 \sim H^{-\alpha}$, as shown in Figure 2, m-q in the main text).

## VII. High-magenetic-field Measurements on SSNO

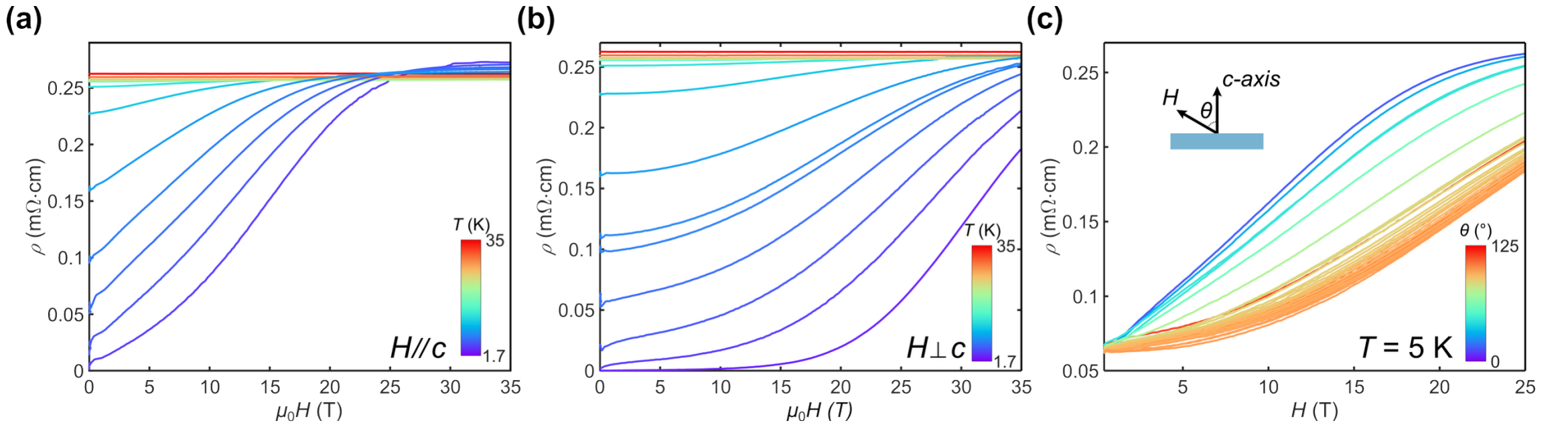

**Figure S7. Resistivity ($\rho$) versus magnetic field up to 35 T on a $Sm_{0.8}Sr_{0.2}NiO_2$ (SSNO) sample** with field applied (a) out of plane, (b) in plane and (c) with a polar rotation angle $\theta$ with respect to the sample surface normal at 5 K.

Magnetoresistance was measured on an SSNO sample with field up to 35 T. Figures S7 shows the data with $H//c$, $H\perp c$, and $H$ applied with a changing polar rotation angle $\theta$ (between the direction of $H$ and the normal to the sample surface).

## VIII. Magnetotransport Measurements on the SCSE0.06 Sample

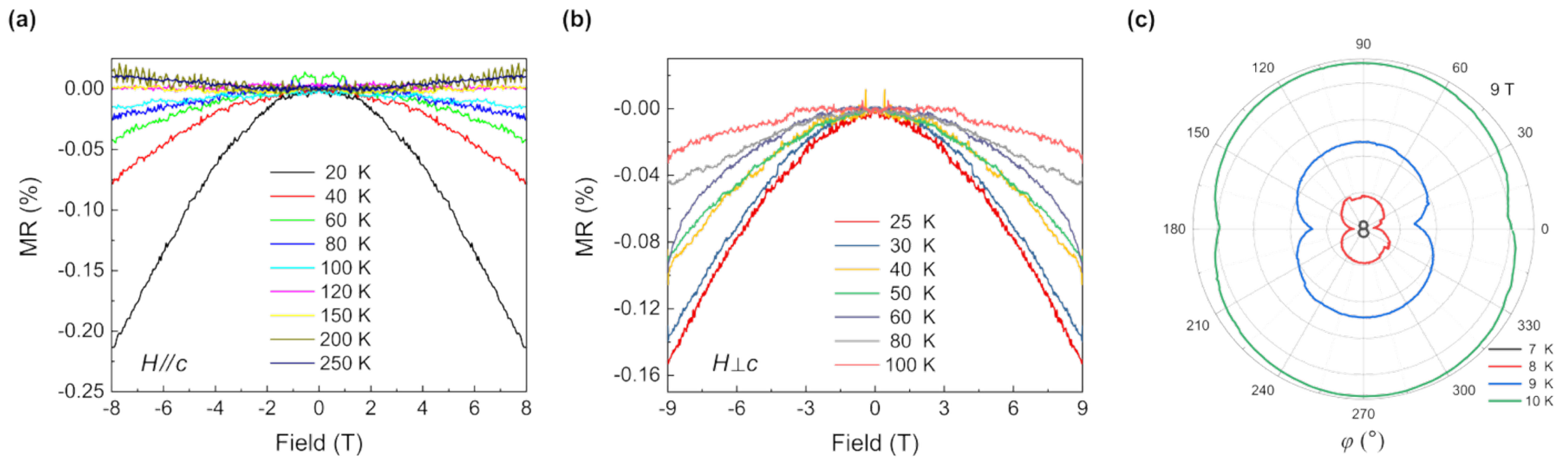

**Figure S8. The normal-state magnetoresistance of the $SCSE_{0.06}$ sample.** (a) and (b) represent the magnetoresistance (MR) under the out-of-plane and in-plane magnetic field at different temperatures. (c) shows the in-plane azimuthal angular dependence of magnetoresistance (with respect to $\varphi$) at different temperatures under a constant magnetic field of 9 T.

The $SCSE_{0.06}$ sample shows a negative normal-state magnetoresistance (MR; symmetrized) under both out-of-plane and in-plane magnetic fields at low temperatures (for out-of-plane field, below 120 K), as illustrated in **Figure S8**, a and b. The strengths of the MR with in-plane and out-of-plane fields are similar. For the in-plane azimuthal angular dependence of magnetoresistance (Figure S7c), a clear twofold symmetry ($C_2$) in the angular dependence is observed, consistent with the results for $(La,Sr)NiO_2$ and $(Pr,Sr)NiO_2$ [3,4].

## IX. Estimation of the Doping Level in the Phase Diagram

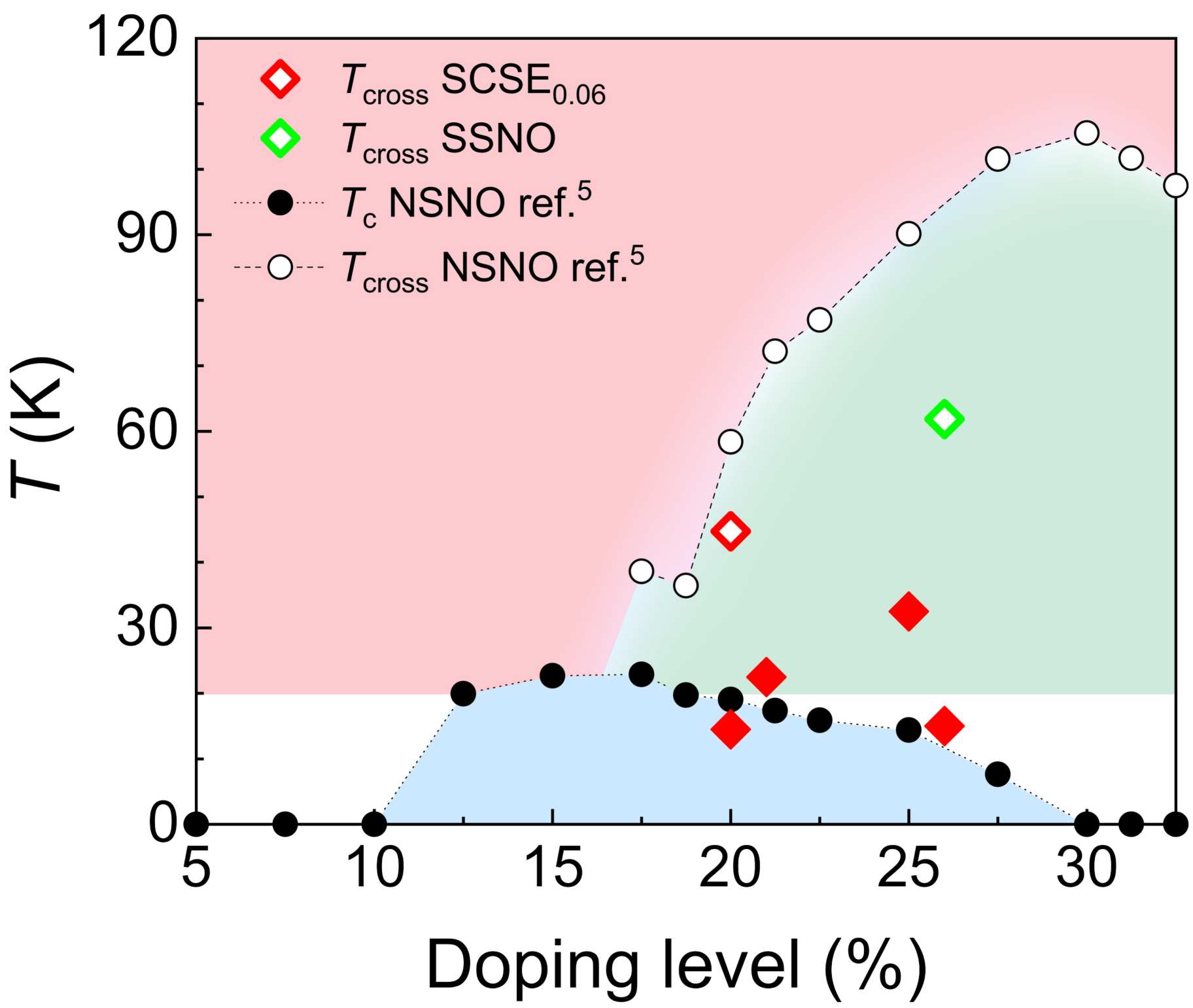

**Figure S9. $T_c$ and $T_{cross}$ versus the 'nominal hole doping level' of the samples in our study plotted in reference to the data extracted from Ref.[5] for $Nd_{1-x}Sr_xNiO_2$ (NSNO).** Here $SCSE_{0.06}$ and SSNO are short for $Sm_{0.74}Ca_{0.01}Sr_{0.19}Eu_{0.06}NiO_2$ and $Sm_{0.8}Sr_{0.2}NiO_2$ respectively.

We plot the $T_c$ and $T_{cross}$ (defined in the main text) versus the 'nominal hole doping level' (defined as the sum of Eu and Ca, assuming both Eu and Ca are +2 dopants) as shown in **Figure S9**. In Figure 4b, the hole doping level is estimated based on the ratio of $Eu^{2+}/Eu^{3+}$ extracted from comparing the distinct ranges of the 'superconducting dome' in Sr [5] and Eu [6] doped systems, therefore the error bars in the main text: the dome size is $0.15 < x < 0.4$ for Eu doping and $0.1 < x < 0.3$ for Sr doping. We therefore estimate that there are 0.05 to 0.1 Eu are $Eu^{3+}$, for which we use the average value of 0.075 to offset and 'calibrate' the hole doping level with 0.025 as the error bar added on Figure 4b.

## References

1. Stankiewicz, J., Bartolomé, J. & Fruchart, D. Spin disorder scattering in magnetic metallic alloys. *Phys. Rev. Lett.* **89**, 106602 (2002).
2. Palstra, T. T. M., Batlogg, B., Schneemeyer, L. F. & Waszczak, J. V. Thermally Activated Dissipation in $B_{2.2}Sr_2Ca_{0.8}Cu_2O_{8+\delta}$. *Phys. Rev. Lett.* **61**, 1662–1665 (1988).
3. Wang, B. Y. *et al.* Effects of rare-earth magnetism on the superconducting upper critical field in infinite-layer nickelates. *Sci. Adv.* **9**, eadf6655 (2023).
4. Ji, H. *et al.* Rotational symmetry breaking in superconducting nickelate $Nd_{0.8}Sr_{0.2}NiO_2$ films. *Nat. Commun.* **14**, 7155 (2023).
5. Lee, K. *et al.* Linear-in-temperature resistivity for optimally superconducting (Nd,Sr)$NiO_2$. *Nature* **619**, 288–292 (2023).
6. Wei, W., Vu, D., Zhang, Z., Walker, F. J. & Ahn, C. H. Superconducting $Nd_{1-x}Eu_xNiO_2$ thin films using in situ synthesis. *Sci. Adv.* **9**, eadh3327 (2023).